\documentclass[12pt]{article}
 \pdfoutput=1
\usepackage{amssymb}
\usepackage{amsmath}
\usepackage{amstext}
\usepackage{graphicx,epsfig}
\usepackage{epsfig}
\usepackage{verbatim} 
\usepackage{fancybox}
\usepackage{color}
\usepackage{ulem}
\usepackage{enumitem}
\usepackage{subfigure}
\usepackage{bbm}
\usepackage{parskip}
\usepackage{cite}
\usepackage{youngtab}
\usepackage{graphicx}
\usepackage{jheppub}
\usepackage{setspace}
\usepackage{longtable}
\usepackage{tabularx}
\usepackage{amsmath,amssymb,color,mathrsfs,verbatim,bbm,wasysym,pstricks,epsfig,colortbl}
\usepackage{slashed,mathtools,youngtab,xcolor,rotating,color,multirow,placeins}

\newcommand{\Comment}[1]{{}}
\definecolor{MyDarkBlue}{rgb}{0.15,0.15,0.45}
\usepackage[linktocpage=true]{hyperref}
\hypersetup{
colorlinks=true,
citecolor=MyDarkBlue,
linkcolor=MyDarkBlue,
urlcolor=MyDarkBlue,
pdfauthor={Kurt Hinterbichler},
pdftitle={},
pdfsubject={hep-th}
}

\newcommand{\be}{\begin{equation}}
\newcommand{\ee}{\end{equation}}
\newcommand{\bea}{\begin{eqnarray}}
\newcommand{\eea}{\end{eqnarray}}
\newcommand{\beas}{\begin{eqnarray*}}
\newcommand{\eeas}{\end{eqnarray*}}
\newcommand{\nn}{\nonumber}

\definecolor{Gray}{gray}{0.9}

\def\({\left(}
\def\){\right)}

\numberwithin{equation}{section}

\begin{document}

\begin{center}
{\LARGE {Partially Massless Higher-Spin Theory II: \\ \vspace{.1in} One-Loop Effective Actions}}
\end{center} 
 \vspace{1truecm}
\thispagestyle{empty} \centerline{
{\large  { Christopher Brust$^{a,}$}}\footnote{E-mail: \Comment{\href{mailto:cbrust@perimeterinstitute.ca}}{\tt cbrust@perimeterinstitute.ca}},
{\large  { Kurt Hinterbichler$^{b,}$}}\footnote{E-mail: \Comment{\href{mailto:kurt.hinterbichler@case.edu}}{\tt kurt.hinterbichler@case.edu}}
                                                          }

\vspace{1cm}

\centerline{{\it 
${}^a$Perimeter Institute for Theoretical Physics,}}
 \centerline{{\it 31 Caroline St. N, Waterloo, Ontario, Canada, N2L 2Y5 }} 
 
 \vspace{.3cm}

\centerline{\it ${}^{\rm b}$CERCA, Department of Physics, Case Western Reserve University, }
\centerline{\it 10900 Euclid Ave, Cleveland, OH 44106, USA}

\begin{abstract}

We continue our study of a generalization of the $D$-dimensional linearized Vasiliev higher-spin equations to include a tower of partially massless (PM) fields. We compute one-loop effective actions by evaluating zeta functions for both the ``minimal'' and ``non-minimal'' parity-even versions of the theory.  Specifically, we compute the log-divergent part of the effective action in odd-dimensional Euclidean AdS spaces for $D=7$ through $19$ (dual to the $a$-type conformal anomaly of the dual boundary theory), and the finite part of the effective action in even-dimensional Euclidean AdS spaces for $D=4$ through $8$ (dual to the free energy on a sphere of the dual boundary theory). We pay special attention to the case $D=4$, where module mixings occur in the dual field theory and subtlety arises in the one-loop computation.  The results provide evidence that the theory is UV complete and one-loop exact, and we conjecture and provide evidence for a map between the inverse Newton's constant of the partially massless higher-spin theory and the number of colors in the dual CFT.

\end{abstract}

\newpage

\tableofcontents
\newpage

\section{Introduction\label{sec:intro}}
\parskip=5pt
\normalsize

In \cite{ustoappear}, we presented evidence for a partially massless higher-spin theory which extends the Vasiliev theory \cite{Vasiliev:1990en, Vasiliev:1992av, Vasiliev:1999ba, Vasiliev:2003ev} (see \cite{Vasiliev:1995dn, Vasiliev:1999ba, Bekaert:2005vh, Iazeolla:2008bp, Didenko:2014dwa,Giombi:2016ejx} for reviews) to include additional partially massless states \cite{Bekaert:2013zya,Basile:2014wua,Grigoriev:2014kpa,Alkalaev:2014nsa,Joung:2015jza}. Furthermore, we presented evidence that the theory is dual to the $\square^2$ CFT which we studied in \cite{Brust:2016gjy} (see also \cite{Osborn:2016bev,Guerrieri:2016whh,Nakayama:2016dby,Peli:2016gio,Gwak:2016sma}). 

On the CFT side, a quantity of interest is the partition function on a sphere of radius $r$. For even CFT dimension $d$, the unambiguous and regulator-independent part of the sphere partition function is the coefficient of the log divergence, corresponding to the $a$-type conformal anomaly $a_{\rm CFT}$,
\be -\ln Z[r]_{\rm CFT}={\rm power\ divergent}+ a_{\rm CFT}\log(r)+{\rm finite},\ \ \ d\ {\rm even},\ee
where the scale of the log and the divergent contributions are set by some UV cutoff.  
For odd CFT dimensions, the unambiguous and regulator independent part is the finite part, known as $F_{\rm CFT}$,
\be -\ln Z[r]_{\rm CFT}= {\rm power\ divergent}+F_{\rm CFT},\ \ \ d\ {\rm odd}.\ee 
For notational convenience, we follow \cite{Giombi:2014xxa} and define a generalized free energy 

\begin{equation}\tilde{F}_{CFT}=\sin\left(\frac{\pi d}{2}\right) \ln Z[r]_{\mathrm{CFT}}\, ,\end{equation}
valid in any $d$, which (up to a constant) reduces to $a_{\rm CFT}$ in even $d$ and $F_{\rm CFT}$ in odd $d$,
\be \tilde{F}_{CFT}=\begin{cases} (-1)^{\frac{d}{2}} \frac{\pi}{2} ~a_{\rm CFT}\, , & d {\rm \ even} \\
(-1)^{\frac{d+1}{2}}~ F_{\rm CFT}\, , & d {\rm \ odd} \end{cases} \, .\ee
In \cite{Brust:2016gjy}, we computed $a_{\rm CFT}$ in even $d$ and the free energy $F_{\rm CFT}$ on spheres in odd $d$ for both the usual $\square$ scalar as well as a $\square^2$ scalar in various dimensions (see also the earlier work \cite{Dowker:2010qy,Dowker:2013oqa,Dowker:2013ysa,Dowker:2015xya,Beccaria:2015uta}).  

On the AdS side, a quantity of interest is the partition function on global Euclidean AdS of radius $R$.  In AdS of odd dimension $D$, the log divergent part $a_{AdS}$ is unambiguous and regulator-independent, and in AdS of even dimension the finite part ${F}_{AdS}$ is regulator-independent and unambiguous.  
\bea &&-\ln Z[R]_{\rm AdS}={\rm power\ divergent}+ a_{AdS}\log(R)+{\rm finite},\ \ \ D\ {\rm odd}  \nn\\
 && -\ln Z[R]_{\rm AdS}={\rm power\ divergent}+{F}_{AdS},\ \ \ D\ {\rm even} 
  \eea
  where the scale of the log and divergent contributions are set by some IR cutoff.
  As in the CFT, we define a generalized free energy $\tilde{F}_{AdS}$ valid in any $D$,
\be \tilde{F}_{AdS}=\begin{cases} (-1)^{\frac{D-1}{2}} \frac{\pi}{2} ~a_{AdS}\, , & D {\rm \ odd} \\
(-1)^{\frac{D}{2}}~F_{AdS} \, ,& D {\rm \ even} \end{cases} \, .\ee
  
  The AdS theory has a perturbative expansion in powers of the $D$-dimensional Newton's constant $G_N$, the dimensionful coupling appearing in front of the action, $S\propto{1\over G_N}\int d^Dx\left(\cdots\right)$.  (We are taking $G_N$ to be dimensionless by implicitly combining it with appropriate powers of the AdS radius $R$, and leaving an overall dimensionless constant multiple appearing in front of it ambiguous.)  Thus $\tilde{F}_{AdS}$ has a perturbative expansion
\begin{equation}\tilde{F}_{AdS} = G_N^{-1}\tilde{F}_0 + \tilde{F}_1 + G_N \tilde{F}_2 + G_N^2 \tilde{F}_3+  \cdots\end{equation}
The lowest part of the expansion, $G_N^{-1}\tilde{F}_0$, is the classical action evaluated on AdS, the next part $\tilde{F}_1$ is the one-loop determinant of the quadratic part of the action expanded on AdS, and the higher parts $\tilde{F}_2,\tilde{F}_3,\cdots$ are higher order bubble diagrams containing the bulk interaction vertices.

AdS/CFT tells us that the well-defined parts of the field theory and AdS partition functions should be equal,
\be \tilde{F}_{CFT}=\tilde{F}_{AdS}.\ee

In the unitary Vasiliev theories, there is an argument that the inverse Newton's constant $G_N^{-1}$ should be quantized \cite{Maldacena:2011jn}.  
Furthermore, we expect on general grounds that $G_N^{-1}\sim N$, where $N$ is the number of ``colors'' of the dual CFT.   In the examples of interest where the dual CFT is free, the generalized free energy in the CFT for the $U(N)$ and $O(N)$ models can be related to the generalized free energy $\tilde{F}$ of a single free real scalar, due to the fact that the CFTs are free and the generalized free energies are additive. Therefore 
\be \tilde{F}_{CFT} = n_{\mathrm{scalars}} \tilde{F},\ \ \ \   n_{\mathrm{scalars}}=\begin{cases} N, & O(N)~\mathrm{ theory} \\ 2N, &  U(N) ~\mathrm{ theory} \end{cases}.\ee
 In AdS, a computation of $\tilde{F}_0$ would require the knowledge of the Vasiliev action, which at present is not universally agreed upon (see \cite{Vasiliev:1988sa,Boulanger:2011dd,Doroud:2011xs,Boulanger:2012bj,Boulanger:2015kfa,Bekaert:2015tva,Bonezzi:2016ttk,Sleight:2016dba} for efforts in this direction). Nevertheless, the answer is expected to be,
  \be \tilde{F}_0= \begin{cases} N\tilde{F}, & {\rm minimal\ Vasiliev \ theory} \\ 2N\tilde{F}, & {\rm non\ minimal\ Vasiliev\ theory} \end{cases}.\ee
   In order to be consistent with the quantization of the inverse Newton's constant, we would expect all the higher corrections $\tilde{F}_2,\tilde{F}_3,\cdots$ to vanish. This however leaves open the possibility that $\tilde{F}_1$ is nontrivial, representing a one-loop renormalization of the inverse Newton's constant.  In order to respect the quantization of the inverse Newton's constant, it must be the case that $\tilde{F}_1$ is an integer multiple of $\tilde{F}$, 
 \be \tilde{F}_1= \begin{cases} n\tilde{F}_1, & {\rm minimal }~ O(N)~\mathrm{ theory} \\ 2n\tilde{F}, & {\rm non\ minimal} ~ U(N) ~\mathrm{ theory} \end{cases},\ \ \ n\in {\rm Integers}, \ee
 so that we may (schematically) move $\tilde{F}_1$ to the left-hand side of the equation $\tilde{F}_{CFT} = G_N^{-1}\tilde{F}_0 + \tilde{F}_1$, giving 
  \be G_N^{-1}\propto N-n.\ee

In the papers by Giombi, Klebanov and Safdi \cite{Giombi:2013fka,Giombi:2014iua} (see also \cite{Jevicki:2014mfa,Beccaria:2014xda,Beccaria:2014zma,Beccaria:2014qea,Giombi:2014yra,Hirano:2015yha,Beccaria:2015vaa,Bae:2016rgm,Pang:2016ofv,Bae:2016hfy,Gunaydin:2016amv,Giombi:2016pvg,Bae:2016xmv,Bae:2017spv}), they did precisely this for the original Vasiliev theory for various $d$, with several different regulators which they demonstrated to be equivalent. They found that in the $U(N)$ theory, $\tilde{F}_1$ vanishes (consistent with $G_N^{-1} \propto N$), and in the $O(N)$ theory $\tilde{F}_1=\tilde{F}$ (consistent with $G_N^{-1}\propto N-1$).

In this paper, we reproduce their computations and perform the analogous computation for the partially massless (PM) theory described in \cite{ustoappear}. We have already computed the conformal anomaly  and free energy in the dual $\square^2$ theory in \cite{Brust:2016gjy}. Both the CFT and AdS theories are not unitary, and so we do not expect the arguments given in \cite{Maldacena:2011jn} for the quantization of the inverse Newton's constant to directly apply\footnote{In particular, they assume the absence of negative-norm states, which the AdS PM theory and its dual have.}. Nevertheless, what we find is the same shift of the inverse Newton's constant as was found for the original Vasiliev theory:
\be G_N^{-1}\propto\begin{cases} N \, ,& {\rm nonminimal/U(N)\ PM\ theory,}\\ N-1 \, ,& {\rm minimal/O(N)\ PM\ theory.} \end{cases}\ee

This is essentially a one-loop computation in the full PM higher-spin theory, and is evidence that the full theory is UV-finite and is a complete theory on its own.
The way we do the computation is to compute a zeta function for the PM theory\footnote{Here, as in the companion paper \cite{ustoappear}, $hs_2$ refers to the algebra of global symmetries of the $\square^2$ CFT, first studied by Eastwood and Leistner \cite{2006math10610E}, then studied further by Joung and Mkrtchyan \cite{Joung:2015jza}.}, $\zeta_{hs_2}(z)$ on $AdS_D$, where $D=d+1$. We evaluate $\zeta_{hs_2}^\prime(0)$, which gives us the one-loop correction $\tilde{F}_1$. We also evaluate $\zeta_{hs_2}(0)$ for even-D spaces, which ought to be 0 so that the log contributions vanish and the finite quantities of interest are unambiguous.  The sum over spins must be regularized in a manner consistent with the higher spin symmetries of the theory, and the authors of \cite{Giombi:2014iua} found that in order to ensure that $\zeta_{hs}(0)=0$ for even-$D$ spaces, one regulator that they could use was to insert $\left(s+\frac{d-3}{2}\right)^{-\alpha}$ in the spin sum, then take the $\alpha\rightarrow 0$ limit afterwards.
One of our findings, identical to the findings of \cite{Gunaydin:2016amv}, is that in order to ensure that $\zeta_{hs_2}(0)=0$ for the PM theory  we need to use that same regulator for the massless particles, and for the partially massless particles we instead need to insert $\left(s+\frac{d-5}{2}\right)^{-\alpha}$, then take the $\alpha\rightarrow 0$ limit afterwards.

The organization of this paper is as follows: in section \ref{sec:generalities}, we define the zeta functions of interest, as well as the spectra of interest, and explain how to extract the one-loop $\tilde{F}_1$ from the zeta function. (Now that we have established that we're only interested in the one-loop effective action, we drop the subscript $1$ to ease the notation, using instead $\tilde{F}^{1-\mathrm{loop}}$ to represent this quantity as needed.) In section \ref{sec:anomalies}, we compute the one-loop renormalization of the inverse Newton's constant in odd $7\leq D \leq 17$, demonstrating that it is consistent with quantization, finding the same results as were obtained for the Vasiliev theory ($G_N^{-1}\propto  N$ for the $U(N)$ theory, $G_N^{-1}\propto  N-1$ for the $O(N)$ theory). Next, in section \ref{sec:freeenergies}, we do the same but in even $D=6,8$, again obtaining matching results.  Finally, we study the case of ${\rm AdS}_{4}$, where the Verma modules of a scalar and a tensor join into one extended Verma module in the dual CFT$_3$, and we are successfully able to compute the zeta function in this case after regularizing the zeta functions of the same two particles, obtaining the same results. 

\textbf {Conventions:} We always use $d$ to refer to the CFT dimension, and $D$ to refer to the AdS dimension, so that $D=d+1$. Despite the fact that we use $d$ below, the computation performed in this paper is exclusively in AdS. $\Delta$ refers to the operator dimension of the CFT dual of an AdS field, and is used to encode the boundary conditions of the AdS field.


\section{Generalities of the One-Loop Renormalization}
\label{sec:generalities}

The one-loop partition function is formally given by
\be Z_{AdS}^{1-\mathrm{loop}}[r]=e^{-W_{AdS}^{1-\mathrm{loop}}}=\prod_{\rm particles}\left(\det {\cal D}\right)^{-\frac{1}{2}}\,, 
\ee
where ${\cal D}$ is the differential operator coming from the quadratic action around AdS of a given particle (with gauge modes appropriately fixed and appropriate Faddeev-Popov ghosts added).  To compute the operator determinants, the zeta function technique is used.

\subsection{Zeta Function Definitions}

For a given theory in $d+1$ dimensions, we would like to compute a regularized total zeta function $\zeta_{d}(z)$, which is related to the one-loop effective action as

\begin{equation}W_{\mathrm{AdS}}^{1-\mathrm{loop}} = \frac{1}{2}\sum_{\mathrm{particles}} \ln\det\mathcal{D} = -\frac{1}{2} \lim_{z\rightarrow 0}\left(\zeta_d(z)\ln(\Lambda^2) + \zeta_d^\prime(z)\right)\end{equation}

\noindent where $\Lambda$ is a UV cutoff in units of the AdS scale, and its coefficient $\zeta_d(0)$ must be zero in order for the physical quantity $\zeta_d^\prime(0)$, which encodes $\tilde{F}^{1-\mathrm{loop}}$, to be unambiguous. $\zeta_d(0)$ being $0$ follows straightforwardly from the definition in odd $D$/even $d$ spaces, but its vanishing is more intricate in even $D$/odd $d$ and must be checked with care.

The total zeta function $\zeta_d(z)$ is given schematically by summing the $\zeta$-function of each particle in the theory,
\be \zeta_{d}(z)=\sum_{\rm particles}\zeta_{d,\Delta,s}(z) .\ee
 We say schematically because this sum is divergent and requires regularization, which we describe below. The zeta function of a single particle can be defined as (see {\cite{Hawking:1976ja,Camporesi:1991nw,Camporesi:1993mz,Camporesi:1994ga,Gaberdiel:2010ar, Gaberdiel:2010xv, Gupta:2012he} for more on the origin of these expressions):

\begin{equation}\zeta_{d,\Delta,s}(z) = \frac{\mathrm{vol}(\mathrm{AdS}_{d+1})}{\mathrm{vol}(S^d)}\frac{2^{d-1}}{\pi}g_{s,d} \int_0^\infty du~ \frac{\mu_{d,s}(u)}{\left(u^2+\left(\Delta-\frac{d}{2}\right)^2\right)^z}\end{equation}

\noindent for $\Delta > \frac{d}{2}$. Zeta functions for $\Delta \leq \frac{d}{2}$ are defined from the above by analytic continuation. Note again here that we use dual CFT notation $d$, $\Delta$, $s$ for convenience of specifying boundary conditions, but the computation is a purely AdS one. The various functions used in this definition are:
\begin{align}\mathrm{vol}(\mathrm{AdS}_{D}) &= \begin{cases}\frac{2(-\pi)^{\frac{D-1}{2}} \log(R)}{\Gamma\left(\frac{D+1}{2}\right)} & D~\mathrm{odd} \\ \pi^{\frac{D-1}{2}} \Gamma\left(-\frac{D-1}{2}\right) & D~\mathrm{even}\end{cases} \, ,\\
\mathrm{vol}(S^d) &= \frac{2 \pi^{\frac{d+1}{2}}}{\Gamma\left(\frac{d+1}{2}\right)}\, , \\
g_{s,d} &= \frac{(2s+d-2)(s+d-3)!}{(d-2)!s!} \, ,\\
\mu_{d~\mathrm{even},s}(u) &= \frac{\pi \left(u^2 + \left(s+\frac{d-2}{2}\right)^2\right)}{\left(2^{d-1}\Gamma\left(\frac{d+1}{2}\right)\right)^2} \prod_{j=0}^{\frac{d-4}{2}}(u^2+j^2) \, ,\\
\mu_{d~\mathrm{odd},s}(u) &= \left(1-\frac{2}{1+e^{2\pi u}}\right)\frac{u \pi  \left(\left(\frac{d-2}{2}+s\right)^2+u^2\right) }{\left(2^{d-1} \Gamma \left(\frac{d+1}{2}\right)\right)^2}\prod _{j=\frac{1}{2}}^{\frac{d-4}{2}} \left(u^2+j^2\right) \, .
\end{align}
The volumes are self-explanatory\footnote{Note, though, that in odd $D$, we have an IR divergence which arises from the infinite volume of AdS. This divergence is the AdS dual of the logarithmic divergence due to the conformal anomaly in the CFT. In even $D$, we have a finite effective action, matching the finite free energy of the CFT.}, $g_{s,d} $ is the number of propagating degrees of freedom in a  massive spin $s$ particle in $d+1$ dimensions, and $\mu$ are spectral densities. We will need Faddeev-Popov-type anticommuting ghosts to eliminate gauge degrees of freedom, and for these the zeta function gets an overall minus sign (i.e. they carry negative degrees of freedom).

The physical quantity of interest, $\tilde{F}^{1-\mathrm{loop}}$, is encoded in the effective action as the linearization of the total $\zeta$-function about $z=0$ \cite{Hawking:1976ja},
\begin{equation}W = -\frac{1}{2} \zeta_{d}^\prime(0) - \zeta_{d}(0) \ln (\Lambda )\, .\end{equation}
Given the vanishing of $\zeta_d(0)$, the contributions to $\tilde F_1$ from each particle are then given in terms of the zeta function as
\begin{align}
a_{d,\Delta,s} &= -\frac{\zeta_{d,\Delta,s}^\prime(0)}{2\log(R)}\, , \qquad \qquad d {\rm \ even}\, , \\
F_{d,\Delta,s} &= -\frac{\zeta_{d,\Delta,s}^\prime(0)}{2}\, ,  \qquad \qquad d {\rm \ odd} \, .
\end{align}
(We will drop the AdS subscripts from now on, as all the remaining computations are performed in the bulk).

Finally, we obtain the full one-loop effective action by summing over all particles in the theory. We must regulate the sum over spins for both the massless and partially massless towers. As stated in the introduction, we will find that for the partially massless tower, the following regularization scheme ensures that $\zeta_d(0)=0$: first we regulate by inserting $\left(s+\frac{d-5}{2}\right)^{-\alpha}$, then perform the sum over $s$, then take the limit $\alpha\rightarrow 0$.

\subsection{The Four Spectra of Interest}

We study four theories in this paper: the nonminimal and minimal Vasiliev theories, and the nonminimal and minimal PM theories. The Vasiliev theories have been studied before in this context \cite{Giombi:2013fka, Giombi:2014iua}, we nevertheless reproduce their work as it is a necessary and natural stepping stone to studying the zeta functions for the PM theories. The PM theories on AdS have fields with wrong sign kinetic terms in the spectrum.  However, since we are computing a functional determinant, the overall normalization and sign of the quadratic action does not matter and these fields still enter the zeta function with a {\it positive} sign.  All the gauge fields, including the partially massless fields, have associated Faddeev-Popov anticommuting ghosts.  These contribute a zeta function of opposite sign.  Therefore, using the results of \cite{ustoappear} and our claimed regularization scheme, the spectrum of each of these four theories and their associated regularized zeta functions are as follows:
1) Nonminimal Vasiliev:
\begin{itemize}
\item $\Delta=d-2$ scalar
\item $\Delta = d+s-2$ physical spins with spin $s$, $s\geq 1$
\item $\Delta = d+s-1$ ghost spins with spin $s-1$, $s \geq 1$
\end{itemize}
\begin{equation}\zeta_{hs,d}^{\mathrm{nonmin}}(z) = \zeta_{d,d-2,0}(z) + \lim_{\alpha \rightarrow 0} \sum_{s=1}^\infty \left(\zeta_{d,d+s-2,s}(z)-\zeta_{d,d+s-1,s-1}(z)\right)\left(s+\frac{d-3}{2}\right)^{-\alpha}\end{equation}
2) Minimal Vasiliev:
\begin{itemize}
\item $\Delta=d-2$ scalar
\item $\Delta = d+s-2$ physical spins with spin $s$, even $s\geq 2$
\item $\Delta = d+s-1$ ghost spins with spin $s-1$, even $s \geq 2$
\end{itemize}
\begin{equation}\zeta_{hs,d}^{\mathrm{min}}(z) = \zeta_{d,d-2,0}(z) + \lim_{\alpha \rightarrow 0} \sum_{s=2,4,6,\ldots}^\infty \left(\zeta_{d,d+s-2,s}(z)-\zeta_{d,d+s-1,s-1}(z)\right)\left(s+\frac{d-3}{2}\right)^{-\alpha}\end{equation}
3) Nonminimal PM:
\begin{itemize}
\item $\Delta=d-2$ scalar
\item $\Delta = d-4$ ``new scalar''
\item $\Delta = d-3$ ``new vector''
\item $\Delta = d-2$ ``new tensor'' ($s=2$)
\item $\Delta = d+s-2$ physical spins with spin $s$, $s\geq 1$
\item $\Delta = d+s-1$ ghost spins with spin $s-1$, $s \geq 1$
\item $\Delta = d+s-4$ PM spins with spin $s$, $s\geq 3$
\item $\Delta = d+s-1$ PM ghost spins with spin $s-3$, $s \geq 3$
\end{itemize}
\begin{align}\zeta_{hs_2,d}^{\mathrm{nonmin}}(z) = &\zeta_{d,d-2,0}(z) + \zeta_{d,d-4,0}(z) + \zeta_{d,d-3,1}(z) + \zeta_{d,d-2,2}(z) \nonumber \\ 
&+\lim_{\alpha \rightarrow 0} \sum_{s=1}^\infty \left(\zeta_{d,d+s-2,s}(z)-\zeta_{d,d+s-1,s-1}(z)\right)\left(s+\frac{d-3}{2}\right)^{-\alpha} \nonumber \\
&+\lim_{\alpha \rightarrow 0} \sum_{s=3}^\infty \left(\zeta_{d,d+s-4,s}(z)-\zeta_{d,d+s-1,s-3}(z)\right)\left(s+\frac{d-5}{2}\right)^{-\alpha}\end{align}
4) Minimal PM:
\begin{itemize}
\item $\Delta=d-2$ scalar
\item $\Delta = d-4$ ``new scalar''
\item $\Delta = d-2$ ``new tensor''($s=2$)
\item $\Delta = d+s-2$ physical spins with spin $s$, even $s\geq 2$
\item $\Delta = d+s-1$ ghost spins with spin $s-1$, even $s \geq 2$
\item $\Delta = d+s-4$ PM spins with spin $s$, even $s\geq 4$
\item $\Delta = d+s-1$ PM ghost spins with spin $s-3$, even $s \geq 4$
\end{itemize}
\begin{align}\zeta_{hs_2,d}^{\mathrm{nonmin}}(z) = &\zeta_{d,d-2,0}(z) + \zeta_{d,d-4,0}(z) +\zeta_{d,d-2,2}(z) \nonumber \\ 
&+\lim_{\alpha \rightarrow 0} \sum_{s=2,4,6,\ldots}^\infty \left(\zeta_{d,d+s-2,s}(z)-\zeta_{d,d+s-1,s-1}(z)\right)\left(s+\frac{d-3}{2}\right)^{-\alpha} \nonumber \\
&+\lim_{\alpha \rightarrow 0} \sum_{s=4,6,8,\ldots}^\infty \left(\zeta_{d,d+s-4,s}(z)-\zeta_{d,d+s-1,s-3}(z)\right)\left(s+\frac{d-5}{2}\right)^{-\alpha}\end{align}

As we will see below in section \ref{sec:freeenergies}, there is a subtlety in $D=4$ for the two PM theories, having to do with module mixing in the dual CFT, requiring modification of the definition of the zeta function. We will give precise definitions there.


\section{One-Loop Renormalization in Odd $D$}
\label{sec:anomalies}

We first turn to the simpler case of computing the zeta function in odd $D$ (even $d$), returning to even $D$ in section \ref{sec:freeenergies}. We will find that the one-loop contribution to the dual of the conformal anomaly equals 0 in the nonminimal theories, and equals the conformal anomaly $a_{CFT}$ of a single real $\square$ and $\square^2$ scalar in the minimal Vasiliev and PM theories, respectively.

We begin with the case of ${\rm AdS}_9$ as an example of the general procedure, because as argued in \cite{Brust:2016gjy,ustoappear}, the cases of $D=3,5,7$ are special for various reasons. We then state results through ${\rm AdS}_{17}$ for completeness' sake. We also study the case of ${\rm AdS}_{7}$, following the n\"aive procedure of simply computing the zeta function, and we encounter no obstacles and obtain the expected result. We do not consider the cases of ${\rm AdS}_{3}$ and ${\rm AdS}_{5}$; we might expect to be able to obtain similar results which match the log theories in ${\rm CFT}_{2}$ and ${\rm CFT}_{4}$. We have not yet performed this check, as the PM theory described in the companion paper \cite{ustoappear} instead produces the duals of the finite ${\rm CFT}_{2}$ and ${\rm CFT}_{4}$ rather than the log theories (see \cite{Brust:2016gjy} for our terminology regarding log vs. finite theories in these special cases).

\subsection{${\rm AdS}_9$}

There are four non-gauge particles in the PM theory which are fully massive, and these must be treated separately. Their representations, given in terms of $(\Delta,s)$, are $(4,0)$, $(5,1)$, $(6,2)$ and $(6,0)$. In the case of $(4,0)$, as $\Delta = \frac{d}{2}$, in order to make the integrals converge we must\footnote{As stated previously, we must do this whenever $\Delta \leq \frac{d}{2}$.} consider $\Delta = 4+\epsilon$ and at the end continue $\epsilon$ to $0$. Upon doing this the zeta function is

\begin{align}\zeta_{8,4+\epsilon,0}(z)=\frac{\epsilon ^{3-2z} \log (R)  \Gamma \left(z-\frac{9}{2}\right)}{215040 \sqrt{\pi } \Gamma (z)} \Bigg(&70 (2 z-9) \epsilon ^4 +49 (2 z-9) (2 z-7) \epsilon ^2 \nonumber \\ &+12 (2 z-9) (2 z-7) (2 z-5)+35 \epsilon ^6\Bigg)\, .\end{align}

Differentiating at $z=0$ produces the contribution to the anomaly, which however starts at $O(\epsilon^3)$,
\begin{equation}a_{8,4+\epsilon,0} = -\frac{\epsilon ^3 \left(-35 \epsilon ^6+630 \epsilon ^4-3087 \epsilon ^2+3780\right)}{12700800}\, .\end{equation}
Therefore
\begin{equation}a_{8,4,0}=\lim_{\epsilon\rightarrow 0} a_{8,4+\epsilon,0} = 0\, .\end{equation}

For the other three non-gauge particles, there are no issues directly computing their zeta functions and evaluating their derivatives: \FloatBarrier
{\renewcommand{\arraystretch}{2} \begin{table}[h]\centering\begin{tabular}{ll}
$\zeta_{8,5,1}(z) = \frac{(8 z (32 z-241)+3507) \log (R) \Gamma \left(z-\frac{9}{2}\right)}{40320 \sqrt{\pi } \Gamma (z-1)}$ & $a_{8,5,1} = -\frac{167}{113400} $ \\
\rowcolor{Gray}$\zeta_{8,6,2}(z) = \frac{2^{-2 z-5} (4 z (25 z-44)-993) \log (R) \Gamma \left(z-\frac{9}{2}\right)}{9 \sqrt{\pi } \Gamma (z-1)}$  & $a_{8,6,2} = \frac{331}{5670}$ \\
$\zeta_{8,6,0}(z) = \frac{2^{-2 z-5} (2 z-7) (6 z+13) \log (R) \Gamma \left(z-\frac{9}{2}\right)}{105 \sqrt{\pi } \Gamma (z-1)}$  & $a_{8,6,0} = \frac{13}{28350}$ \end{tabular} \end{table} } \FloatBarrier

For the gauge fields, we must sum over each tower of spins and each tower of corresponding ghosts. We give one example here then state answers for the other cases of interest. For the spin sums, we follow the procedure of \cite{Giombi:2014iua}  and perform the sum over spins {\it before} performing the $u$-integral in the definition of the zeta function. We define the $u$-integrand of the zeta function simply by $\zeta(z,u)$, 
\begin{equation}\zeta_{8,\mathrm{spins}}(z) = \int_0^\infty du~ \sum_{s=1}^\infty \zeta_{8,6+s,s}(z,u)\,.\end{equation}
The result of performing the sum is
\begin{align}\zeta_{8,\mathrm{spins}}(z,u)=&\frac{u^2 2^{-2 z-9} \log (R) \left(u^2+1\right)^{-z}}{14175 \pi } \nonumber \\ & \times \Bigg(4 \Big(4^z (8 \zeta (2 (z-5))+26 \zeta (2 (z-4))-28 \zeta (2 (z-3))
-6 \zeta (2 (z-2))+\zeta (2 z-11) \nonumber \\ &+23 \zeta (2 z-9)-\zeta (2 z-7)-23 \zeta (2 z-5))-2880 \left(u^2+1\right) \left(4 u^2+1\right) \left(4 u^2+9\right)\Big)
 \nonumber \\ &+u^2 4^z \Big(8 \left(7 u^2+5\right) \zeta (2 (z-6))-4 \left(9 u^2+35\right) \zeta (2 (z-4))+\left(35-78 u^2\right) \zeta (2 z-9) \nonumber \\ &+\left(6 u^4+8 u^2\right) \zeta (2 (z-7))-2 \left(3 u^4+14 u^2-77\right) \zeta (2 (z-5))+\left(u^4+u^2\right) \zeta (2 z-15) \nonumber \\ &+\left(10 u^4+28 u^2+5\right) \zeta (2 z-13)+\left(-11 u^4+49 u^2+119\right) \zeta (2 z-11)-54 \zeta (2 (z-3)) \nonumber \\ &-159 \zeta (2 z-7)\Big)\Bigg)\, .\end{align}
The result of integrating this with respect to $u$ is
\begin{align}\zeta_{8,\mathrm{spins}}(z)=&\frac{4^{-z-7} \log (R)\Gamma \left(z-\frac{9}{2}\right)}{14175 \sqrt{\pi } \Gamma (z)} \times \Bigg(15\ 2^{2 z+1} (8 z-15) \zeta (2 (z-7))+15\ 4^{z+2} z (2 z-9) \zeta (2 (z-6))\nonumber \\ 
&+2^{2 z+3} (z (z (32 z-105)-809)+2772) \zeta (2 (z-5)) -69120 (z-1) (2 z-7) (6 z+13)\nonumber \\ 
&-2^{2 z+1} (2 z-9) (2 z-7) (112 z-199) \zeta (2 (z-3))-4^z (2 z-9) (2 z-7) (8 z+457) \zeta (2 z-7)\nonumber \\ 
&+15\ 2^{2 z+1} (z-1) \zeta (2 z-15)+15\ 4^z (4 z (z+6)-119) \zeta (2 z-13)\nonumber \\ 
&+4^z (2 z (2 z (8 z+273)-4405)+13461) \zeta (2 z-11)-3\ 2^{2 z+3} (2 z-9) (2 z-7) (2 z-5) \zeta (2 z-4)\nonumber \\ 
&+4^z (2 z-9) (2 z (184 z-999)+1315) \zeta (2 z-9)+2^{2 z+3} (2 z-9) (z (52 z-417)+755) \zeta (2 z-8)\nonumber \\ 
&-23\ 4^{z+1} (2 z-9) (2 z-7) (2 z-5) \zeta (2 z-5)\Bigg) \,. \end{align}
Finally, we may differentiate at 0 to obtain
\begin{equation}a_{8,\mathrm{spins}}=-\frac{\zeta_{8,\mathrm{spins}}^\prime(0)}{2\log(R)}=-\frac{14334496157}{31261590360000}\, .\end{equation}

These same steps may be performed for the other sums of interest. The corresponding zeta functions and contributions to the anomaly are \FloatBarrier
{\renewcommand{\arraystretch}{2} \begin{table}[h]\centering\begin{tabular}{rclrcl}
$\zeta_{8,\mathrm{ghosts}}(z)$ & $ = $ & $\displaystyle \int_0^\infty du~ \sum_{s=1}^\infty \zeta_{8,7+s,s-1}(z,u)$ & $ a_{8,\mathrm{ghosts}}$&$=$ & $ \frac{624643}{31261590360000}$ \\
\rowcolor{Gray}$\zeta_{8,\mathrm{even~spins}}(z)$ & $ = $ & $\displaystyle  \int_0^\infty du~ \sum_{s=2,4,6,\ldots}^\infty \zeta_{8,6+s,s}(z,u) $ & $a_{8,\mathrm{even~spins}} $&$= $ & $ -\frac{22329082757 }{62523180720000} $\\
$\zeta_{8,\mathrm{even~ghosts}}(z)$ & $ = $ & $\displaystyle  \int_0^\infty du~ \sum_{s=2,4,6,\ldots}^\infty \zeta_{8,7+s,s-1}(z,u) $ & $a_{8,\mathrm{even~ghosts}} $&$=$ & $  -\frac{6339909557}{62523180720000} $\\
\rowcolor{Gray}$\zeta_{8,\mathrm{PM~spins}}(z)$ & $ =$ & $\displaystyle   \int_0^\infty du~ \sum_{s=3}^\infty \zeta_{8,4+s,s}(z,u) $ & $ a_{8,\mathrm{PM}} $&$=$ & $  -\frac{1778854645457 }{31261590360000}$\\
$\zeta_{8,\mathrm{PM~ghosts}}(z)$ & $ = $ & $\displaystyle  \int_0^\infty du~ \sum_{s=3}^\infty \zeta_{8,7+s,s-3}(z,u) $ & $ a_{8,\mathrm{PM~ghosts}} $&$=$ & $ \frac{78710743 }{31261590360000}$\\
\rowcolor{Gray}$\zeta_{8,\mathrm{even~PM~spins}}(z)$ & $ = $ & $\displaystyle  \int_0^\infty du~ \sum_{s=4,6,8,\ldots}^\infty \zeta_{8,4+s,s}(z,u) $ & $ a_{8,\mathrm{even~PM}} $&$=$ & $  -\frac{3684874361057}{62523180720000}$\\
$\zeta_{8,\mathrm{even~PM~ghosts}}(z)$ & $ =$ & $\displaystyle   \int_0^\infty du~ \sum_{s=4,6,8,\ldots}^\infty \zeta_{8,7+s,s-3}(z,u)$ & $ a_{8,\mathrm{even~PM~ghosts}} $&$=$ & $  \frac{35089486543}{62523180720000}$ \end{tabular} \end{table} } \FloatBarrier

With all of these results, we may now sum up and compare with the CFT for each of the theories of interest. First we reproduce the results of \cite{Giombi:2014iua} for the nonminimal original Vasiliev theory, 
\begin{equation}a_{hs,8}^{\mathrm{nonmin}}=a_{8,6,0} + a_{8,\mathrm{spins}} - a_{8,\mathrm{ghosts}} = 0\, .\end{equation}
\noindent Therefore $G_N^{-1}\propto  N$.

Now the minimal original Vasiliev theory theory,
\begin{equation}a_{hs,8}^{\mathrm{min}}=a_{8,6,0} + a_{8,\mathrm{even~spins}} - a_{8,\mathrm{even~ghosts}} = \frac{23}{113400}\, .\end{equation}
This is precisely the anomaly of one real $\square$ scalar in 8d. Therefore we may interpret $G_N^{-1}\propto  N-1$, as in \cite{Giombi:2014iua}.

Now the PM theory. We begin with the nonminimal theory,
\begin{align}a_{hs_2,8}^{\mathrm{nonmin}} = &a_{8,6,0} + a_{8,\mathrm{spins}} - a_{8,\mathrm{ghosts}} \nonumber \\
&+a_{8,4,0} + a_{8,5,1} + a_{8,6,2} + a_{8,\mathrm{PM~spins}} - a_{8,\mathrm{PM~ghosts}} = 0\, .\end{align}
This is consistent with $G_N^{-1}\propto N$, with no one-loop correction.

Finally, the minimal PM theory:
\begin{align}a_{hs_2,8}^{\mathrm{min}} = &a_{8,6,0} + a_{8,\mathrm{even~spins}} - a_{8,\mathrm{even~ghosts}} \nonumber \\
&+a_{8,4,0} + a_{8,6,2} + a_{8,\mathrm{even~PM~spins}} - a_{8,\mathrm{even~PM~ghosts}} = -\frac{13}{14175}\,.\end{align}
This is precisely the conformal anomaly of one real $\square^2$ scalar in 8d, which supports the interpretation $G_N^{-1}\propto  N-1$.

\subsection{${\rm AdS}_7$}

In ${\rm AdS}_7$, the only expected subtlety comes from the two scalars, whose dual CFT modules mix \cite{Brust:2016gjy}. Indeed, the free action for the scalars is nondiagonalizable \cite{ustoappear}. However, following the n\"aive procedure of simply computing the zeta function seems to give us the expected results.  In the future, it would be interesting to inquire as to why this happens.

The only subtlety in ${\rm AdS}_7$ is the fact that the $\Delta = 2$ scalar and the $\Delta = 3$ vector have $\Delta \leq \frac{d}{2}$, and so their contributions require analytic continuation from $\Delta > \frac{d}{2}$. Computing the zeta function for a scalar of dimension $\Delta$ and continuing, we obtain

\begin{equation}\zeta^\prime_{6,\Delta,0}(0)=-\frac{(\Delta -3)^3 \left(3 \Delta ^4-36 \Delta ^3+141 \Delta ^2-198 \Delta +82\right)}{7560} \log R\end{equation}

As $\Delta \rightarrow 2 $, we obtain $a_{6,2,0} = -\frac{1}{1512}$. Similarly for the $\Delta = 3$ vector, we obtain $a_{6,3,1}=0$.

The rest of the computation follows similarly to the ${\rm AdS}_9$ case above. In the end, we obtain the following results: \FloatBarrier
{\renewcommand{\arraystretch}{2} \begin{table}[h]\centering\begin{tabular}{rclrcl}
$a_{6,4,0}$ & $ = $ & $ \frac{1}{1512}$ & $a_{6,2,0}$ & $ = $ & $ -\frac{1}{1512}$ \\
\rowcolor{Gray}$a_{6,3,1}$ & $ = $ & $ 0$ & $a_{6,4,2}$ & $ = $ & $ \frac{109}{1890}$ \\
$a_{6,\mathrm{spins}}$ &$=$ &$-\frac{1124261}{1702701000}$  & $a_{6,\mathrm{ghosts}}$ &$=$ &$\frac{233}{212837625}$  \\
\rowcolor{Gray}$a_{6,\mathrm{even~spins}}$ &$=$ &$-\frac{1125659}{851350500}$ &$a_{6,\mathrm{even~ghosts}}$ &$=$ &$\frac{1127057}{1702701000}$  \\
$a_{6,\mathrm{PM}}$ &$=$ &$-\frac{98159381}{1702701000}$  & $a_{6,\mathrm{PM~ghosts}}$ &$=$ &$-\frac{543703}{851350500}$  \\
\rowcolor{Gray}$a_{6,\mathrm{even~PM}}$ &$=$ &$-\frac{89282353}{1702701000}$ &$a_{6,\mathrm{even~PM~ghosts}}$ &$=$ &$-\frac{2219257}{425675250}$
\end{tabular} \end{table} } \FloatBarrier
\begin{equation}a_{hs,6}^{\mathrm{nonmin}}=a_{6,4,0} + a_{6,\mathrm{spins}} - a_{6,\mathrm{ghosts}} = 0\, .\end{equation}
\begin{equation}a_{hs,6}^{\mathrm{min}}=a_{6,4,0} + a_{6,\mathrm{even~spins}} - a_{6,\mathrm{even~ghosts}} = -\frac{1}{756}\, .\end{equation}
\begin{align}a_{hs_2,6}^{\mathrm{nonmin}} = &a_{6,4,0} + a_{6,\mathrm{spins}} - a_{6,\mathrm{ghosts}} \nonumber \\
&+a_{6,2,0} + a_{6,3,1} + a_{6,4,2} + a_{6,\mathrm{PM~spins}} - a_{6,\mathrm{PM~ghosts}} = 0\, .\end{align}
\begin{align}a_{hs_2,6}^{\mathrm{min}} = &a_{6,4,0} + a_{6,\mathrm{even~spins}} - a_{6,\mathrm{even~ghosts}} \nonumber \\
&+a_{6,2,0} + a_{6,4,2} + a_{6,\mathrm{even~PM~spins}} - a_{6,\mathrm{even~PM~ghosts}} = \frac{8}{945}\,.\end{align}
These results all support the conclusion that $G_N^{-1} \propto N$ in the nonminimal Vasiliev and PM theories, and $G_N^{-1} \propto N-1$ in the minimal Vasiliev and PM theories.

\subsection{${\rm AdS}_{11}$ Through ${\rm AdS}_{17}$}

Carrying out the above procedure in ${\rm AdS}_{11}$ through ${\rm AdS}_{17}$, we fill out the following tables of contributions to $a$.
The contributions of the four massive particles are given in table \ref{tab:massivecontributions}. The spin sums, their associated ghosts' sums, and the difference between them (we've  included the difference for convenience) are in table \ref{tab:spinanomalies}.  The same spin sums, but with even spins only, are in table \ref{tab:evenanomalies}. The sum over the partially massless particles and their associated ghosts is in table \ref{tab:pmanomalies}, and finally, the same but with even spins only is in table \ref{tab:evenpmanomalies}. \FloatBarrier

{\renewcommand{\arraystretch}{1.5}
\begin{table}[h]
\centering
\begin{tabular}{|c|c|c|c|c|}\hline
$D$ & $a_{\mathrm{scalar}}$ & $a_{\mathrm{new~scalar}}$ & $a_{\mathrm{new~vector}}$& $a_{\mathrm{new~tensor}}$ \\ \hline
\rowcolor{Gray} 7 & $\frac{1}{1512}$ & $-\frac{1}{1512}$ & $0$ & $\frac{109}{1890}$\\
9 & $-\frac{13}{14175}$ & 0 & $\frac{167}{56700}$ & $-\frac{331}{2835}$ \\
\rowcolor{Gray} 11 & $-\frac{19}{30800}$ & $-\frac{263}{7484400}$ & $\frac{1049}{467775}$ & $-\frac{243}{2200}$ \\
13 & $-\frac{275216}{638512875}$ & $-\frac{28151}{1277025750}$ & $\frac{22419}{14014000}$ & $-\frac{9492016}{91216125}$ \\
\rowcolor{Gray} 15 &$-\frac{307525}{980755776}$ & $-\frac{717}{56056000}$ & $\frac{2229232}{1915538625}$ & $-\frac{12075925}{122594472}$ \\
17 & $-\frac{70327}{297797500}$ & $-\frac{531926}{69780335625}$ & $\frac{3964165}{4547140416}$ & $-\frac{797931}{8508500}$ \\ \hline
\end{tabular}
\caption{The one-loop contributions of the massive particles to the dual of the conformal anomaly in ${\rm AdS}_{7}$ through ${\rm AdS}_{17}$.}
\label{tab:massivecontributions}
\end{table}
}

{\renewcommand{\arraystretch}{1.5}
\begin{table}[h]
\centering
\begin{tabular}{|c|c|c|c|}\hline
$d$  & $a_{\mathrm{spins}}$ & $a_{\mathrm{ghosts}}$ & $a_{\mathrm{difference}}$\\ \hline
\rowcolor{Gray} 7 & $ -\frac{1124261}{1702701000}$ & $ \frac{233}{212837625}$ & $ -\frac{1}{1512}$ \\
9 & $\frac{14334496157}{15630795180000}$ & $-\frac{624643}{15630795180000}$ & $\frac{13}{14175}$\\
\rowcolor{Gray} 11 & $\frac{19887362021}{32238515058750}$  & $-\frac{269057}{257908120470000}$ & $\frac{19}{30800}$ \\
13  & $\frac{19659148636669746041}{45610068020048532000000}$ & $-\frac{1509998285959}{45610068020048532000000}$ & $\frac{275216}{638512875}$  \\
\rowcolor{Gray} 15 & $\frac{2937757532570636610049}{9369068139118302615000000}$ &  $-\frac{5570293999663}{4684534069559151307500000}$ & $\frac{307525}{980755776}$  \\
17 & $\frac{517155640022646178755331547867}{2189879516259542026449129600000000}$ & $-\frac{101884121512763172133}{2189879516259542026449129600000000}$ &  $\frac{70327}{297797500}$  \\ \hline
\end{tabular}
\caption{The one-loop contribution of the massless spins, their ghosts, and the difference of the two to the dual of the conformal anomaly in ${\rm AdS}_{7}$ through ${\rm AdS}_{17}$.}
\label{tab:spinanomalies}
\end{table}
}

{\renewcommand{\arraystretch}{1.5}
\begin{table}[h]
\centering
\begin{tabular}{|c|c|c|c|c|}\hline
$d$  & $a_{\mathrm{even~spins}}$ &  $a_{\mathrm{even~ghosts}}$ & $a_{\mathrm{difference}}$ \\ \hline
\rowcolor{Gray} 7 & $ -\frac{1125659}{851350500}$ & $ \frac{1127057}{1702701000}$ & $ -\frac{1}{504}$ \\
9 &  $\frac{22329082757}{31261590360000}$ & $\frac{6339909557}{31261590360000}$ & $\frac{29}{56700}$\\
\rowcolor{Gray} 11 &  $\frac{336323718943}{515816240940000}$ &  $-\frac{18125926607}{515816240940000}$ & $\frac{5143}{7484400}$ \\
13  & $\frac{38721009127060464041}{91220136040097064000000}$ & 
$\frac{597288146279028041}{91220136040097064000000}$& $\frac{1423223}{3405402000}$ \\
\rowcolor{Gray} 15 & $\frac{2949756676401053999087}{9369068139118302615000000}$ & $-\frac{5999571915208694519}{4684534069559151307500000}$ & $\frac{38754643}{122594472000}$ \\
17 & $\frac{1033175411772321536794365707867}{4379759032519084052898259200000000}$ & $\frac{1135868272970820716297387867}{4379759032519084052898259200000000}$ & $\frac{7366432081}{31261590360000}$ \\ \hline
\end{tabular}
\caption{The one-loop contribution of the even massless spins, their ghosts, and the difference of the two to the dual of the conformal anomaly in ${\rm AdS}_{7}$ through ${\rm AdS}_{17}$.}
\label{tab:evenanomalies}
\end{table}
}

{\renewcommand{\arraystretch}{1.5}
\begin{table}[h]
\centering
\begin{tabular}{|c|c|c|c|c|c|c|c|}\hline
$d$ & $a_{\mathrm{PM~spins}}$&  $a_{\mathrm{PM~ghosts}}$  & $a_{\mathrm{difference}}$ \\ \hline
\rowcolor{Gray} 7 & $ -\frac{98159381}{1702701000}$ & $ -\frac{543703}{851350500}$ & $ -\frac{431}{7560}$ \\
9 & $\frac{1778854645457}{15630795180000}$ & $-\frac{78710743}{15630795180000}$ & $\frac{239}{2100}$\\
\rowcolor{Gray} 11 & $\frac{7426137840569443}{68603560045020000}$  & $-\frac{2294093807}{68603560045020000}$ & $\frac{54011}{498960}$\\ 
13 & $\frac{3454885655909454389459}{33711789406122828000000}$  & $-\frac{19136712972541}{33711789406122828000000}$ & $\frac{1046987549}{10216206000}$ \\
\rowcolor{Gray} 15 & $\frac{10945175018472155430073063}{112428817669419631380000000}$ & $-\frac{1540451871354437}{112428817669419631380000000}$ & $\frac{442030453}{4540536000}$ \\ 
17 & $\frac{29067924098063852463799436333081}{312839930894220289492732800000000}$ & $-\frac{126297330828409506919}{312839930894220289492732800000000}$ & $\frac{322745647937}{3473510040000}$\\ \hline
\end{tabular}
\caption{The one-loop contribution of the partially massless spins, their ghosts, and the difference of the two to the dual of the conformal anomaly in ${\rm AdS}_{7}$ through ${\rm AdS}_{17}$.}
\label{tab:pmanomalies}
\end{table}
}

{\renewcommand{\arraystretch}{1.5}
\begin{table}[h]
\centering
\begin{tabular}{|c|c|c|c|c|c|c|c|}\hline
$d$ & $a_{\mathrm{even~PM~spins}}$ & $a_{\mathrm{even~PM~ghosts}}$ & $a_{\mathrm{difference}}$\\ \hline
\rowcolor{Gray} 7 & $ -\frac{89282353}{1702701000}$ & $ -\frac{2219257}{425675250}$ & $ -\frac{17}{360}$ \\
9 & $\frac{3684874361057}{31261590360000}$ & $-\frac{35089486543}{31261590360000}$ & $\frac{2249}{18900}$ \\
\rowcolor{Gray} 11 & $\frac{15136962033791593}{137207120090040000}$ & $\frac{23004885601693}{137207120090040000}$  & $\frac{3569}{32400}$ \\ 
13 & $\frac{7019560472352046241459}{67423578812245656000000}$ & $-\frac{1927792373316186541}{67423578812245656000000}$  & $\frac{27279877}{261954000}$\\
\rowcolor{Gray} 15 & $\frac{22150846580406974501675563}{224857635338839262760000000}$ & $\frac{1184338144028746310563}{224857635338839262760000000}$  & $\frac{4025400551}{40864824000}$\\ 
17 & $\frac{58681950547410701097873345293081}{625679861788440578985465600000000}$ & $-\frac{639052947373902172972626919}{625679861788440578985465600000000}$  & $\frac{57490751477}{612972360000}$\\ \hline
\end{tabular}
\caption{The one-loop contribution of the even partially massless spins, their ghosts, and the difference of the two to the dual of the conformal anomaly in ${\rm AdS}_{7}$ through ${\rm AdS}_{17}$.}
\label{tab:evenpmanomalies}
\end{table}
}

\FloatBarrier

Putting these results all together, we obtain the results for the one-loop correction to the inverse Newton's constant in all four of these theories in table \ref{tab:totalresults}. \FloatBarrier

{\renewcommand{\arraystretch}{1.5}
\begin{table}[h]
\centering
\begin{tabular}{|c|c|c|c|c|}\hline
$d$ & Nonmin Vasiliev & Min Vasiliev & Nonmin PM & Min PM  \\ \hline
\rowcolor{Gray} 7 & 0 & $-\frac{1}{756} $ &0 & $ \frac{8}{945}$\\
9 & 0 & $\frac{23}{113400}$ & 0 & $-\frac{13}{14175}$\\
\rowcolor{Gray} 11 & 0 & $-\frac{263}{7484400}$ & 0 & $\frac{62}{467775}$ \\ 
13 & 0 & $\frac{133787}{20432412000}$ & 0 & $-\frac{28151}{1277025750}$ \\
\rowcolor{Gray} 15 & 0 & $-\frac{157009}{122594472000}$ & 0 & $\frac{7636}{1915538625}$ \\
17 & 0 & $\frac{16215071}{62523180720000}$ & 0 & $-\frac{1488889}{1953849397500}$ \\ \hline
\end{tabular}
\caption{Complete result for the AdS computation of anomalies at one loop.}
\label{tab:totalresults}
\end{table}
}

\FloatBarrier


\section{One-Loop Renormalization in Even $D$}
\label{sec:freeenergies}

In even-$D$ cases, we must not only concern ourselves with the finite part of the effective action (which will be dual to the free energy $F$), but also with the log-divergent part of the action, the would-be $a$-type conformal anomaly. Odd-dimensional CFTs have no $a$-type conformal anomaly due to the absence of diff-invariant counterterms to renormalize the log divergence, and so our regularization scheme for the AdS dual of the free energy must guarantee that there is no dual log divergence as well, in the process ensuring that the free energy is unambiguous and physical. In terms of zeta functions, the free energy will be manifested in terms of $\zeta_d^\prime(0)$, whereas the log divergence will be $\zeta_d(0)$. We calculate these two independently but with the same regulator.

What we will find is that the idea behind the regulator of \cite{Giombi:2014iua}, inserting $\left(s+\frac{d-3}{2}\right)^{-\alpha}$ before carrying out the spin sum, may continue to be used for the partially massless tower, but needs to be modified to $\left(s+\frac{d-5}{2}\right)^{-\alpha}$ (as found also in \cite{Gunaydin:2016amv}). The massless regulator is left unchanged. Note that in this section, we subtract ghosts from spins {\it before} regulating and performing the spin sums. Therefore, in all results below, when we say ``spins'', what we really mean is ``spins minus ghosts''.

The one-loop computation in even $D$ is much more technically involved than the odd $D$ computation. To that end, we need to define some helpful intermediate functions, following \cite{Giombi:2014iua}. 
First, the spectral density contains a term $1-\frac{2}{1+e^{2\pi u}}$. We define two partial spectral densities by splitting up this term:
\begin{equation}\mu^{(1)} = \frac{u \pi  \left(\left(\frac{d-2}{2}+s\right)^2+u^2\right) }{\left(2^{d-1} \Gamma \left(\frac{d+1}{2}\right)\right)^2}\prod _{j=\frac{1}{2}}^{\frac{d-4}{2}} \left(u^2+j^2\right)\, ,\end{equation}
\begin{equation}\mu^{(2)} = -\frac{2 u \pi  \left(\left(\frac{d-2}{2}+s\right)^2+u^2\right) }{\left(e^{2 \pi  u}+1\right) \left(2^{d-1} \Gamma \left(\frac{d+1}{2}\right)\right)^2}\prod _{j=\frac{1}{2}}^{\frac{d-4}{2}} \left(u^2+j^2\right)\, .\end{equation}
We use these to define partial zeta functions:
\begin{equation}\zeta^{(i)}_{d,\Delta,s}(z) = \frac{\mathrm{vol}(\mathrm{AdS}_{d+1})}{\mathrm{vol}(S^d)}\frac{2^{d-1}}{\pi}g_{s,d} \int_0^\infty du~ \frac{\mu^{(i)}_{d,s}(u)}{\left(u^2+\left(\Delta-\frac{d}{2}\right)^2\right)^z}\, ,\end{equation}
which sum to the (complete) zeta function
\begin{equation}\zeta_{d,\Delta,s}(z) = \zeta^{(1)}_{d,\Delta,s}(z) + \zeta^{(2)}_{d,\Delta,s}(z)\, .\end{equation}
We continue to use the notation $\zeta_{d,\Delta,s}(z,u)$ for the pre-integrated zeta function,
\begin{equation}\zeta_{d,\Delta,s}(z) = \int_0^\infty du~\zeta_{d,\Delta,s}(z,u)\,. \end{equation}
We also need the following helpful identities and definitions:
\begin{equation}\lim_{z\rightarrow 0} \frac{d}{dz}\left(\int_0^\infty du \frac{u^{2p+1}}{\left(u^2+\left(\Delta-\frac{d}{2}\right)^2\right)^z}\right) = (-1)^{p+1}\left(\Delta-\frac{d}{2}\right)^{2(1+p)} \frac{H_{1+p}-2\ln\left(\Delta-\frac{d}{2}\right)}{2(1+p)}\, ,\end{equation}
\noindent where $H_n$ is the $n^\mathrm{th}$ harmonic number. This identity covers all of the single particle $\zeta^{(1)\prime}(0)$ that we need to evaluate. 

We now turn to $\zeta^{(2)\prime}(0)$.  Define the following,
\begin{equation}\int_0^\infty du \frac{u^{2p+1}\ln\left(u^2+\left(\Delta-\frac{d}{2}\right)^2\right)}{1+e^{2\pi u}} = c_p+ 2\int_0^{\Delta-\frac{d}{2}} dx~x A_p(x)\, ,\label{eqn:uint}\end{equation}
where
\begin{align}c_p=\frac{\Gamma(2+2p)}{4^{1+2p}\pi^{2(1+p)}}\Big(&\zeta(2+2p)\left(-2^{1+2p}\ln(2\pi)+\ln(4\pi)+(2^{1+2p}-1)\psi\left(2+2p\right)\right) \nonumber \\
&\qquad +(2^{1+2p}-1)\zeta^\prime (2+2p)\Big)\, , \end{align}
\begin{align}A_p(x)&=\frac{4^p-2}{(4\pi)^{2p}}\Gamma(2p)\zeta(2p)-x^2A_{p-1}(x)\, ,\nonumber \\
A_0(x)&= \frac{1}{2}\psi\left(x+\frac{1}{2}\right)-\frac{1}{2}\ln x\, ,\end{align}
where $\psi$ is the digamma function, and $A_p$ is defined recursively.

We will split the computation of $\zeta^\prime_{d,\Delta,s}(0)$ into two parts, which we will call the ``$J$'' and ``$K$'' parts, following \cite{Giombi:2014iua}. The definitions of these revolve around the $x$-integral that will be done over the polygamma function $\psi\left(x+\frac{1}{2}\right)$. $J$ is the part of the answer that follows by ignoring this integral:
\begin{equation}J_{d,\Delta,s} = \left\{\zeta^\prime_{d,\Delta,s}(0) \mathrel{}\middle|\mathrel{} \psi\left(x+\frac{1}{2}\right)\rightarrow 0\right\}\, .\end{equation}

Then, in terms of this, $K$ is the remaining part of the zeta function, which now only involves the integral of the polygamma function:
\begin{equation}K_{d,\Delta,s} = \zeta^\prime_{d,\Delta,s}(0) -J_{d,\Delta,s}\, .\end{equation}
As mentioned earlier, there are subtleties in $D=4$, which we will explore below.

\subsection{${\rm AdS}_{6}$}

As we will demonstrate, in all four theories we study, $\zeta_5(0)=0$. In the nonminimal theories, we find $\zeta_5^\prime(0)=0$, consistent with $G_N^{-1}\propto N$, and for the minimal theories, we find $\zeta_5^\prime(0)=-2F$, where $F$ is the free energy of a real scalar with an appropriate number of powers of the Laplacian evaluated on $S^5$.

\subsubsection{$\zeta_5(0)$}

$\zeta_5(0)$ receives contributions from every field and ghost in the theory. First we begin with the four massive particles. We may define $\zeta_{5,\Delta,s}(0)$ per particle by integrating then setting $z\rightarrow 0$ for $\zeta^{(1)}$, and the opposite for $\zeta^{(2)}$:
\begin{equation}\zeta_{d,\Delta,s}(0) \equiv \zeta^{(1)}_{d,\Delta,s}(0) + \left(\int_0^\infty du~\zeta^{(2)}_{d,\Delta,s}(0,u)\right)\, .\end{equation}
Carrying this out for the four massive particles we obtain
\begin{align}\zeta_{5,3,0}(0)&=\frac{1}{1512}\, , \qquad \qquad  \zeta_{5,1,0}(0)=-\frac{37}{7560}\, ,\nonumber \\
\zeta_{5,2,1}(0)&=\frac{67}{7560}\, , \qquad \qquad  \zeta_{5,3,2}(0)=\frac{13}{270}\,.\end{align}

The zeta functions for massless and PM spins and their associated ghosts may be done in an identical fashion. After that, we must sum over spins, but again this sum is divergent and must be regulated by inserting a $\left(s+\frac{d-3}{2}\right)^{-\alpha}$ for massless spins or a $\left(s+\frac{d-5}{2}\right)^{-\alpha}$ for PM spins, doing the sum, and then setting $\alpha \rightarrow 0$\footnote{We could ask what would happen if we had instead chosen to regulate the PM sum in $d=5$ by $(s+x)^{-\alpha}$, for some other $x$. If we had done so, we would have found instead:
\begin{align}\zeta_{hs_2,5}^{\mathrm{nonmin,}x}(0) &= \zeta_{5,3,0}(0) + \zeta_{5,1,0}(0) + \zeta_{5,2,1}(0) + \zeta_{5,3,2}(0) \nonumber \\ 
&~+\lim_{\alpha \rightarrow 0} \sum_{s=1}^\infty \left(\zeta_{5,s+3,s}(z)-\zeta_{5,s+4,s-1}(z)\right)\left(s+1\right)^{-\alpha} \nonumber \\
&~+\lim_{\alpha \rightarrow 0} \sum_{s=3}^\infty \left(\zeta_{5,s+1,s}(z)-\zeta_{5,s+4,s-3}(z)\right)\left(s+x\right)^{-\alpha} \nonumber \\
&=\frac{x \left(105 x^8-1050 x^6+3423 x^4-4510 x^2+1480\right)}{151200}\end{align}
Thus we see that we ought to choose $x=0$ to ensure that the above vanishes. We can also carry out this same exercise for the minimal theory, and in other dimensions. We have done so and all support the conclusion that the appropriate regulator is $\left(s+\frac{d-5}{2}\right)^{-\alpha}$.
}:
\begin{equation}\zeta_{5,\mathrm{spins}}(0)=\lim_{\alpha\rightarrow 0}\sum_{s=1}^\infty \left(\zeta_{5,s+1,s}(0)-\zeta_{5,s+2,s-1}(0)\right)\left(s+\frac{d-3}{2}\right)^{-\alpha}=-\frac{1}{1512}\, ,\end{equation}
\begin{equation}\zeta_{5,\mathrm{PM~spins}}(0)=\lim_{\alpha\rightarrow 0}\sum_{s=3}^\infty \left(\zeta_{5,s-1,s}(0)-\zeta_{5,s+2,s-3}(0)\right)\left(s+\frac{d-5}{2}\right)^{-\alpha}=-\frac{197}{3780}\, .\end{equation}
In the case of even spins only:
\begin{equation}\zeta_{5,\mathrm{even~spins}}(0)=\lim_{\alpha\rightarrow 0}\sum_{s=2,4,6,\ldots}^\infty \left(\zeta_{5,s+1,s}(0)-\zeta_{5,s+2,s-1}(0)\right)\left(s+\frac{d-3}{2}\right)^{-\alpha}=-\frac{1}{1512}\, .\end{equation}
\begin{equation}\zeta_{5,\mathrm{even~PM~spins}}(0)=\lim_{\alpha\rightarrow 0}\sum_{s=4,6,8,\ldots}^\infty \left(\zeta_{5,s-1,s}(0)-\zeta_{5,s+2,s-3}(0)\right)\left(s+\frac{d-5}{2}\right)^{-\alpha}=-\frac{109}{2520}\, .\end{equation}
By adding together the appropriate $\zeta_5(0)$s, we see that this regularization scheme is sufficient to ensure that $\zeta_{hs,5}^\mathrm{nonmin}(0)$, $\zeta_{hs,5}^{\mathrm{min}}(0)$, $\zeta_{hs_2,5}^{\mathrm{nonmin}}(0)$, and $\zeta_{hs_2,5}^{\mathrm{min}}(0)$ are all $0$, thus there is no dual conformal anomaly term for the ${\rm CFT}_5$.

\subsubsection{$\zeta_5^\prime(0)$}

As in \cite{Giombi:2014iua}, the computation of $\zeta^\prime_5(0)$ is considerably more involved. We generally refer to the procedure outlined there, with modifications as needed to accommodate the PM theory. We split all of the computations into ``$J$'' and ``$K$'' pieces, as explained above. We begin with $J$. It receives contributions from both $\zeta^{(1)\prime}(0)$ and $\zeta^{(2)\prime}(0)$. We begin with the computation of $\zeta^{(1)\prime}(0)$. This may be evaluated as in \cite{Giombi:2014iua} by using the identities defined above.  Now, we turn to $\zeta^{(2)\prime}(0)$.

\begin{equation}\zeta^{(2)\prime}_{5,\Delta,s}(0)=-\int_0^\infty du~\frac{(s+1) (s+2) (2 s+3) u \left(u^2+\frac{1}{4}\right) \left(\left(s+\frac{3}{2}\right)^2+u^2\right) \ln \left(\left(\Delta -\frac{5}{2}\right)^2+u^2\right)}{360 \left(e^{2 \pi  u}+1\right)}\, .\end{equation}

We may expand this in powers of $u$, then use \eqref{eqn:uint} term-by-term to replace each $u$ integral with a constant plus an $x$ integral. After recursing in $p$, we're ultimately left with an $x$ integral of the form $\int_0^{\Delta-\frac{d}{2}}dx~x^q \psi\left(x+\frac{1}{2}\right)$. All such integrals (along with their multiplicative coefficients out front) define what we mean by $K$. Everything else in $\zeta^{(2)\prime}(0)$, along with all of $\zeta^{(1)\prime}(0)$, together define $J$. More details can be found in \cite{Giombi:2014iua}.

All of the $J$ pieces are straightforward to deal with with the identities above. The $K$ pieces require some more work; we defer the reader to the methodology in \cite{Giombi:2013fka, Giombi:2014iua}. The general idea is to rewrite the polygamma function in an integral form
\begin{equation}\psi(y)=\int_0^\infty dt\left(\frac{e^{-t}}{t}-\frac{e^{-yt}}{1-e^{-t}}\right)\, ,\end{equation}
then perform the $x$ integral, then perform the regulated spin sum, subtract off the power-law divergences in the $t$-integral, then finally perform the $t$ integral. We perform the $t$ integral by taking appropriate derivatives so that we can use the integral representation of the Hurwitz-Lerch $\Phi$ function,
\begin{equation}\Phi(z,s,v)=\frac{1}{\Gamma(s)}\int_0^\infty dt~\frac{t^{s-1}e^{-vt}}{1-ze^{-t}}=\sum_{n=0}^\infty \frac{z^n}{(n+v)^{s}}\, ,\end{equation}
\noindent which we can relate in turn to derivatives of the Hurwitz zeta function.

Once all of the dust settles, we find the following results. First, the individual particles:\\
\FloatBarrier
\footnotesize
{\renewcommand{\arraystretch}{1.5}
\begin{table}[!htbp]
\centering
\begin{tabular}{rcl}
$J_{5,\mathrm{scalar}}$&$=$&$\frac{3 \log (A)}{640}-\frac{7 \zeta '(4)}{256 \pi ^4}-\frac{31 \zeta '(6)}{512 \pi ^6}-\frac{1459}{907200}+\frac{89 \gamma }{241920}-\frac{11 \log (2)}{161280}+\frac{89 \log (\pi )}{241920}$ \nonumber \\
\rowcolor{Gray}$K_{5,\mathrm{scalar}}$&$=$&$\frac{23 \log (A)}{1920}+\frac{21\zeta '(-5)}{640} -\frac{7\zeta '(-3)}{192} -\frac{\zeta (3)}{96 \pi ^2}-\frac{\zeta (5)}{32 \pi ^4}-\frac{1181}{1382400}+\frac{211 \log (2)}{483840}$  \nonumber \\
$J_{5,\mathrm{new~scalar}}$&$=$&$\frac{3 \log (A)}{640}-\frac{7 \zeta '(4)}{256 \pi ^4}-\frac{31 \zeta '(6)}{512 \pi ^6}-\frac{4483}{907200}+\frac{89 \gamma }{241920}-\frac{11 \log (2)}{161280}+\frac{89 \log (\pi )}{241920}$ \nonumber \\
\rowcolor{Gray}$K_{5,\mathrm{new~scalar}}$&$=$&$-\frac{99 \log (A)}{640}+\frac{21\zeta '(-5)}{640} +\frac{19\zeta '(-3)}{64} +\frac{3 \zeta (5)}{32 \pi ^4}-\frac{3 \zeta (3)}{32 \pi ^2}+\frac{1433}{51200}+\frac{211 \log (2)}{483840} $\nonumber \\
$J_{5,\mathrm{new~vector}}$&$=$&$\frac{25 \log (A)}{384}-\frac{91 \zeta '(4)}{256 \pi ^4}-\frac{155 \zeta '(6)}{512 \pi ^6}-\frac{737}{36288}+\frac{1033 \gamma }{241920}-\frac{31 \log (2)}{17920}+\frac{1033 \log (\pi )}{241920}$ \nonumber \\
\rowcolor{Gray}$K_{5,\mathrm{new~vector}}$&$=$&$\frac{71 \log (A)}{384}+\frac{21\zeta '(-5)}{128} -\frac{155\zeta '(-3)}{192} +\frac{17 \zeta (3)}{96 \pi ^2}+\frac{5 \zeta (5)}{32 \pi ^4}-\frac{3821}{276480}+\frac{2903 \log (2)}{483840} $\nonumber \\
$J_{5,\mathrm{new~tensor}}$&$=$&$\frac{343 \log (A)}{960}-\frac{245 \zeta '(4)}{128 \pi ^4}-\frac{217 \zeta '(6)}{256 \pi ^6}-\frac{439}{4050}+\frac{383 \gamma }{17280}-\frac{41 \log (2)}{3840}+\frac{383 \log (\pi )}{17280} $\nonumber \\
\rowcolor{Gray}$K_{5,\mathrm{new~tensor}}$&$=$&$\frac{1001 \log (A)}{960}+\frac{147\zeta '(-5)}{320} -\frac{469\zeta '(-3)}{96} -\frac{49 \zeta (3)}{48 \pi ^2}-\frac{7 \zeta (5)}{16 \pi ^4}-\frac{54467}{691200}+\frac{227 \log (2)}{6912}$\end{tabular}\end{table}} 
\normalsize 

\FloatBarrier\noindent where $A$ is Glaisher's constant. Now, the various spin sums:
\footnotesize
{\renewcommand{\arraystretch}{1.5}
\begin{table}[!htbp]
\centering
\begin{tabular}{rcl}
$J_{5,\mathrm{spins}}$&$=$&$-\frac{3 \log (A)}{640}+\frac{7 \zeta '(4)}{256 \pi ^4}+\frac{31 \zeta '(6)}{512 \pi ^6}-\frac{89 \gamma }{241920}+\frac{1459}{907200}+\frac{11 \log (2)}{161280}-\frac{89 \log (\pi )}{241920} $\nonumber \\
\rowcolor{Gray}$K_{5,\mathrm{spins}}$&$=$&$ -\frac{23 \log (A)}{1920}-\frac{21\zeta '(-5)}{640} +\frac{7\zeta '(-3)}{192} +\frac{\zeta (3)}{96 \pi ^2}+\frac{\zeta (5)}{32 \pi ^4}+\frac{1181}{1382400}-\frac{211 \log (2)}{483840} $\nonumber \\
$J_{5,\mathrm{even~spins}}$&$=$&$-\frac{3 \log (A)}{640}+\frac{7 \zeta '(4)}{256 \pi ^4}+\frac{31 \zeta '(6)}{512 \pi ^6}-\frac{89 \gamma }{241920}+\frac{1459}{907200}+\frac{11 \log (2)}{161280}-\frac{89 \log (\pi )}{241920} $\nonumber \\
\rowcolor{Gray}$K_{5,\mathrm{even~spins}}$&$=$&$ -\frac{23 \log (A)}{1920}-\frac{21\zeta '(-5)}{640} +\frac{7\zeta '(-3)}{192} +\frac{5 \zeta (3)}{192 \pi ^2}-\frac{11 \zeta (5)}{128 \pi ^4}+\frac{1181}{1382400}+\frac{7349 \log (2)}{483840} $\nonumber \\
$J_{5,\mathrm{PM~spins}}$&$=$&$ -\frac{41 \log (A)}{96}+\frac{147 \zeta '(4)}{64 \pi ^4}+\frac{155 \zeta '(6)}{128 \pi ^6}-\frac{1621 \gamma }{60480}+\frac{30311}{226800}+\frac{503 \log (2)}{40320}-\frac{1621 \log (\pi )}{60480} $\nonumber \\
\rowcolor{Gray}$K_{5,\mathrm{PM~spins}}$&$=$&$ -\frac{103 \log (A)}{96}-\frac{21\zeta '(-5)}{32} +\frac{259\zeta '(-3)}{48} +\frac{15 \zeta (3)}{16 \pi ^2}+\frac{3 \zeta (5)}{16 \pi ^4}+\frac{22337}{345600}-\frac{4751 \log (2)}{120960} $\nonumber \\
$J_{5,\mathrm{even~PM~spins}}$&$=$&$ -\frac{139 \log (A)}{384}+\frac{497 \zeta '(4)}{256 \pi ^4}+\frac{465 \zeta '(6)}{512 \pi ^6}-\frac{1817 \gamma }{80640}+\frac{34273}{302400}+\frac{1733 \log (2)}{161280}-\frac{1817 \log (\pi )}{80640} $\nonumber \\
\rowcolor{Gray}$K_{5,\mathrm{even~PM~spins}}$&$=$&$ -\frac{341 \log (A)}{384}-\frac{63\zeta '(-5)}{128} +\frac{881\zeta '(-3)}{192} +\frac{289 \zeta (3)}{192 \pi ^2}+\frac{29 \zeta (5)}{128 \pi ^4}+\frac{70243}{1382400}-\frac{14389 \log (2)}{53760}$\end{tabular}\end{table}} 
\normalsize \FloatBarrier

Adding together the appropriate $J$ and $K$ for our four theories gives the claimed results. For nonminimal Vasiliev theory:
\begin{align}-\frac{1}{2}\zeta^{\mathrm{nonmin}\prime}_{hs,5}(0) &= -\frac{1}{2}\Big(J_{5,\mathrm{scalar}}+K_{5,\mathrm{scalar}}+J_{5,\mathrm{spins}}+K_{5,\mathrm{spins}}\Big) \nonumber \\
&=0\, ,\end{align}
and for the minimal Vasiliev theory:
\begin{align}-\frac{1}{2}\zeta^{\mathrm{min}\prime}_{hs,5}(0) &= -\frac{1}{2}\Big(J_{5,\mathrm{scalar}}+K_{5,\mathrm{scalar}}+J_{5,\mathrm{even~spins}}+K_{5,\mathrm{even~spins}}\Big) \nonumber \\
&=\frac{15 \zeta (5)}{256 \pi ^4}-\frac{\zeta (3)}{128 \pi ^2}-\frac{\log (4)}{256}\, .\end{align}
This is the free energy of a real $\square$ scalar on $S^5$. 

Now, the nonminimal PM theory:
\begin{align}-\frac{1}{2}\zeta^{\mathrm{nonmin}\prime}_{hs_2,5}(0) &= -\frac{1}{2}\Big(J_{5,\mathrm{scalar}}+K_{5,\mathrm{scalar}}+J_{5,\mathrm{new~scalar}}+K_{5,\mathrm{new~scalar}}+J_{5,\mathrm{new~vector}}+K_{5,\mathrm{new~vector}} \nonumber \\
&\qquad \qquad +J_{5,\mathrm{new~tensor}}+K_{5,\mathrm{new~tensor}}+J_{5,\mathrm{spins}}+K_{5,\mathrm{spins}}+J_{5,\mathrm{PM~spins}}+K_{5,\mathrm{PM~spins}}\Big) \, ,\nonumber \\
&=0\, ,\end{align}
and the minimal PM theory:
\begin{align}-\frac{1}{2}\zeta^{\mathrm{min}\prime}_{hs_2,5}(0) &= -\frac{1}{2}\Big(J_{5,\mathrm{scalar}}+K_{5,\mathrm{scalar}}+J_{5,\mathrm{new~scalar}}+K_{5,\mathrm{new~scalar}}+J_{5,\mathrm{new~tensor}}+K_{5,\mathrm{new~tensor}} \nonumber \\
&\qquad \qquad +J_{5,\mathrm{even~spins}}+K_{5,\mathrm{even~spins}}+J_{5,\mathrm{even~PM~spins}}+K_{5,\mathrm{even~PM~spins}}\Big) \nonumber \\
&=\frac{15 \zeta (5)}{128 \pi ^4}-\frac{13 \zeta (3)}{64 \pi ^2}+\frac{7 \log (2)}{64}\, .\end{align}
As we demonstrated in \cite{Brust:2016gjy}, this is the free energy of a real $\square^2$ scalar on $S^5$.

\subsection{${\rm AdS}_8$}

The techniques we use for ${\rm AdS}_8$ are identical to the techniques we use for ${\rm AdS}_{6}$, so we simply state results:
{\renewcommand{\arraystretch}{1.5}
\begin{table}[h] \centering \begin{tabular}{rclrcl}
$\zeta_{7,5,0}(0)$&$=$&$\frac{127}{226800}$ & $\zeta_{7,3,0}(0)$&$=$&$-\frac{23}{226800}$ \nonumber \\
\rowcolor{Gray}$\zeta_{7,4,1}(0)$&$=$&$-\frac{311}{226800}$& $\zeta_{7,5,2}(0)$&$=$&$\frac{71}{1200}$ \nonumber \\
$\zeta_{7,\mathrm{spins}}(0)$&$=$&$-\frac{127}{226800}$ & $\zeta_{7,\mathrm{even~spins}}(0)$&$=$&$-\frac{127}{226800}$ \nonumber \\
\rowcolor{Gray}$\zeta_{7,\mathrm{PM~spins}}(0)$&$=$&$-\frac{2617}{45360}$ & $\zeta_{7,\mathrm{even~PM~spins}}(0)$&$=$&$-\frac{3349}{56700}$\end{tabular}\end{table}
} \\
These all sum together to ensure that $\zeta_7(0)=0$ for all four theories. Now $\zeta_7^\prime(0)$: \FloatBarrier
{\renewcommand{\arraystretch}{1.5}
\begin{table}[!htbp]
\centering
\begin{tabular}{rcl}
$J_{7,\mathrm{scalar}}$&$=$&\scriptsize $-\frac{5 \log (A)}{7168}+\frac{259 \zeta '(4)}{61440 \pi ^4}+\frac{155 \zeta '(6)}{12288 \pi ^6}+\frac{127 \zeta '(8)}{8192 \pi ^8}+\frac{139583}{217728000}-\frac{14359 \gamma }{232243200}+\frac{19 \log (2)}{5529600}-\frac{14359 \log (\pi )}{232243200}$\nonumber \\
\rowcolor{Gray} $K_{7,\mathrm{scalar}}$&$=$&\scriptsize $\frac{537 \log (A)}{35840}+\frac{17 \zeta '(-7)}{21504}+\frac{61 \zeta '(-5)}{5120}+\frac{13 \zeta '(-3)}{3072}+\frac{3 \zeta (7)}{128 \pi ^6}-\frac{\zeta (3)}{160 \pi ^2}-\frac{\zeta (5)}{64 \pi ^4}-\frac{2171077}{722534400}-\frac{15157 \log (2)}{232243200}$\nonumber \\
$J_{7,\mathrm{new~scalar}}$&$=$&\scriptsize $ -\frac{5 \log (A)}{7168}+\frac{259 \zeta '(4)}{61440 \pi ^4}+\frac{155 \zeta '(6)}{12288 \pi ^6}+\frac{127 \zeta '(8)}{8192 \pi ^8}-\frac{14359 \gamma }{232243200}+\frac{391481}{1524096000}+\frac{19 \log (2)}{5529600}-\frac{14359 \log (\pi )}{232243200}$\nonumber \\
\rowcolor{Gray} $K_{7,\mathrm{new~scalar}}$&$=$&\scriptsize $-\frac{181 \log (A)}{107520}+\frac{17 \zeta '(-7)}{21504}+\frac{13 \zeta '(-3)}{3072}-\frac{73 \zeta '(-5)}{15360}-\frac{\zeta (3)}{720 \pi ^2}-\frac{\zeta (5)}{192 \pi ^4}-\frac{\zeta (7)}{128 \pi ^6}+\frac{755987}{6502809600}-\frac{15157 \log (2)}{232243200}$\nonumber \\
$J_{7,\mathrm{new~vector}}$&$=$&\scriptsize $-\frac{49 \log (A)}{5120}+\frac{3493 \zeta '(4)}{61440 \pi ^4}+\frac{1829 \zeta '(6)}{12288 \pi ^6}+\frac{889 \zeta '(8)}{8192 \pi ^8}+\frac{741641}{217728000}-\frac{185953 \gamma }{232243200}+\frac{3571 \log (2)}{38707200}-\frac{185953 \log (\pi )}{232243200}$\nonumber \\
\rowcolor{Gray} $K_{7,\mathrm{new~vector}}$&$=$&\scriptsize $-\frac{73 \log (A)}{3072}+\frac{17 \zeta '(-7)}{3072}+\frac{203 \zeta '(-3)}{3072}-\frac{203 \zeta '(-5)}{3072}+\frac{29 \zeta (3)}{1440 \pi ^2}+\frac{13 \zeta (5)}{192 \pi ^4}+\frac{7 \zeta (7)}{128 \pi ^6}+\frac{1549619}{928972800}-\frac{207379 \log (2)}{232243200}$\nonumber \\
$J_{7,\mathrm{new~tensor}}$&$=$&\scriptsize $-\frac{2187 \log (A)}{35840}+\frac{7371 \zeta '(4)}{20480 \pi ^4}+\frac{3627 \zeta '(6)}{4096 \pi ^6}+\frac{3429 \zeta '(8)}{8192 \pi ^8}+\frac{498223}{8064000}-\frac{42839 \gamma }{8601600}+\frac{1013 \log (2)}{1433600}-\frac{42839 \log (\pi )}{8601600}$\nonumber \\
\rowcolor{Gray} $K_{7,\mathrm{new~tensor}}$&$=$&\scriptsize $\frac{12879 \log (A)}{7168}+\frac{153 \zeta '(-7)}{7168}+\frac{27 \zeta '(-5)}{1024}-\frac{2619 \zeta '(-3)}{1024}+\frac{27 \zeta (5)}{64 \pi ^4}+\frac{81 \zeta (7)}{128 \pi ^6}-\frac{81 \zeta (3)}{80 \pi ^2}-\frac{26735727}{80281600}-\frac{48917 \log (2)}{8601600}$\end{tabular}\end{table}} 
\FloatBarrier
Now the spin sums:
{\renewcommand{\arraystretch}{1.5}
\begin{table}[!htbp]
\centering
\begin{tabular}{rcl}
$J_{7,\mathrm{spins}}$&$=$&\scriptsize $\frac{5 \log (A)}{7168}-\frac{259 \zeta '(4)}{61440 \pi ^4}-\frac{155 \zeta '(6)}{12288 \pi ^6}-\frac{127 \zeta '(8)}{8192 \pi ^8}-\frac{139583}{217728000}+\frac{14359 \gamma }{232243200}-\frac{19 \log (2)}{5529600}+\frac{14359 \log (\pi )}{232243200}$\nonumber \\
\rowcolor{Gray} $K_{7,\mathrm{spins}}$&$=$&\scriptsize $-\frac{537 \log (A)}{35840}-\frac{13 \zeta '(-3)}{3072}-\frac{61 \zeta '(-5)}{5120}-\frac{17 \zeta '(-7)}{21504}+\frac{\zeta (3)}{160 \pi ^2}+\frac{\zeta (5)}{64 \pi ^4}-\frac{3 \zeta (7)}{128 \pi ^6}+\frac{2171077}{722534400}+\frac{15157 \log (2)}{232243200}$\nonumber \\
$J_{7,\mathrm{even~spins}}$&$=$&\scriptsize $\frac{5 \log (A)}{7168}-\frac{259 \zeta '(4)}{61440 \pi ^4}-\frac{155 \zeta '(6)}{12288 \pi ^6}-\frac{127 \zeta '(8)}{8192 \pi ^8}-\frac{139583}{217728000}+\frac{14359 \gamma }{232243200}-\frac{19 \log (2)}{5529600}+\frac{14359 \log (\pi )}{232243200}$\nonumber \\
\rowcolor{Gray} $K_{7,\mathrm{even~spins}}$&$=$&\scriptsize $-\frac{537 \log (A)}{35840}-\frac{13 \zeta '(-3)}{3072}-\frac{61 \zeta '(-5)}{5120}-\frac{17 \zeta '(-7)}{21504}+\frac{11 \zeta (3)}{3072 \pi ^2}+\frac{21 \zeta (5)}{1024 \pi ^4}+\frac{15 \zeta (7)}{2048 \pi ^6}+\frac{2171077}{722534400}-\frac{438443 \log (2)}{232243200}$\nonumber \\
$J_{7,\mathrm{PM~spins}}$&$=$&\scriptsize $\frac{73 \log (A)}{1024}-\frac{5173 \zeta '(4)}{12288 \pi ^4}-\frac{12865 \zeta '(6)}{12288 \pi ^6}-\frac{4445 \zeta '(8)}{8192 \pi ^8}-\frac{19949423}{304819200}+\frac{271393 \gamma }{46448640}-\frac{6211 \log (2)}{7741440}+\frac{271393 \log (\pi )}{46448640}$\nonumber \\
\rowcolor{Gray} $K_{7,\mathrm{PM~spins}}$&$=$&\scriptsize $-\frac{9069 \log (A)}{5120}+\frac{683 \zeta '(-5)}{15360}+\frac{2547 \zeta '(-3)}{1024}-\frac{85 \zeta '(-7)}{3072}+\frac{159 \zeta (3)}{160 \pi ^2}-\frac{31 \zeta (5)}{64 \pi ^4}-\frac{87 \zeta (7)}{128 \pi ^6}+\frac{2153990567}{6502809600}+\frac{308659 \log (2)}{46448640}$\nonumber \\
$J_{7,\mathrm{even~PM~spins}}$&$=$&\scriptsize $ \frac{79 \log (A)}{1280}-\frac{5593 \zeta '(4)}{15360 \pi ^4}-\frac{2759 \zeta '(6)}{3072 \pi ^6}-\frac{889 \zeta '(8)}{2048 \pi ^8}-\frac{23638907}{381024000}+\frac{292753 \gamma }{58060800}-\frac{6871 \log (2)}{9676800}+\frac{292753 \log (\pi )}{58060800}$\nonumber \\
\rowcolor{Gray} $K_{7,\mathrm{even~PM~spins}}$&$=$&\scriptsize $-\frac{6893 \log (A)}{3840}-\frac{17\zeta '(-7)}{768} +\frac{1961 \zeta '(-3)}{768}-\frac{83 \zeta '(-5)}{3840}+\frac{47317 \zeta (3)}{46080 \pi ^2}-\frac{1625 \zeta (5)}{3072 \pi ^4}-\frac{1217 \zeta (7)}{2048 \pi ^6}+\frac{21648379}{65028096}+\frac{1127779 \log (2)}{58060800}$\end{tabular}\end{table}} 
\FloatBarrier
Now, we put these ingredients together. First the nonminimal Vasiliev theory:
\begin{align}-\frac{1}{2}\zeta^{\mathrm{nonmin}\prime}_{hs,7}(0) &= -\frac{1}{2}\Big(J_{7,\mathrm{scalar}}+K_{7,\mathrm{scalar}}+J_{7,\mathrm{spins}}+K_{7,\mathrm{spins}}\Big) \nonumber \\
&=0\, . \end{align}
Then the minimal Vasiliev theory:
\begin{align}-\frac{1}{2}\zeta^{\mathrm{min}\prime}_{hs,7}(0) &= -\frac{1}{2}\Big(J_{7,\mathrm{scalar}}+K_{7,\mathrm{scalar}}+J_{7,\mathrm{even~spins}}+K_{7,\mathrm{even~spins}}\Big) \nonumber \\
&=\frac{41 \zeta (3)}{30720 \pi ^2}-\frac{5 \zeta (5)}{2048 \pi ^4}-\frac{63 \zeta (7)}{4096 \pi ^6}+\frac{\log (2)}{1024}\, .\end{align}
This is the free energy of a real $\square$ scalar on $S^7$.  Next the nonminimal PM theory:
\begin{align}-\frac{1}{2}\zeta^{\mathrm{nonmin}\prime}_{hs_2,7}(0) &= -\frac{1}{2}\Big(J_{7,\mathrm{scalar}}+K_{7,\mathrm{scalar}}+J_{7,\mathrm{new~scalar}}+K_{7,\mathrm{new~scalar}}+J_{7,\mathrm{new~vector}}+K_{7,\mathrm{new~vector}} \nonumber \\
&\qquad \qquad +J_{7,\mathrm{new~tensor}}+K_{7,\mathrm{new~tensor}}+J_{7,\mathrm{spins}}+K_{7,\mathrm{spins}}+J_{7,\mathrm{PM~spins}}+K_{7,\mathrm{PM~spins}}\Big) \nonumber \\
&=0\, ,\end{align}
and finally, the minimal PM theory:
\begin{align}-\frac{1}{2}\zeta^{\mathrm{min}\prime}_{hs_2,7}(0) &= -\frac{1}{2}\Big(J_{7,\mathrm{scalar}}+K_{7,\mathrm{scalar}}+J_{7,\mathrm{new~scalar}}+K_{7,\mathrm{new~scalar}}+J_{7,\mathrm{new~tensor}}+K_{7,\mathrm{new~tensor}} \nonumber \\
&\qquad \qquad +J_{7,\mathrm{even~spins}}+K_{7,\mathrm{even~spins}}+J_{7,\mathrm{even~PM~spins}}+K_{7,\mathrm{even~PM~spins}}\Big) \nonumber \\
&=\frac{55 \zeta (5)}{1024 \pi ^4}-\frac{79 \zeta (3)}{15360 \pi ^2}-\frac{63 \zeta (7)}{2048 \pi ^6}-\frac{1}{512} 3 \log (2)\, .\end{align}
As we demonstrated in \cite{Brust:2016gjy}, this is the free energy of a real $\square^2$ scalar on $S^7$.

\subsection{${\rm AdS}_{4}$}

As we demonstrated in \cite{Brust:2016gjy,ustoappear}, the ${\rm AdS}_{4}$/${\rm CFT}_{3}$ PM theory is special because of the new scalar-new tensor module mixing that takes place in both AdS and in the CFT. Therefore, we might expect there to be subtlety in the zeta function for this theory. However, a similar module mixing took place in the ${\rm AdS}_{7}$/${\rm CFT}_{6}$ PM theory between the two scalars, but nothing prevented us from directly computing the zeta function in that case. We might therefore expect that the ${\rm AdS}_4$ case would also be straightforward. However, it is not; the $\zeta^\prime_3(0)$ for both the new scalar and new tensor are ill-defined due to a new divergence that arises. We can, however, regulate both of these divergences by increasing their masses via increasing the scaling dimension of the dual operator by $\epsilon$ for both. As we will see, using the same $\epsilon$ for both is crucial for obtaining the right dual free energy. Upon doing so, we will see that the divergence cancels in the total zeta function, and we obtain the expected results. We do not have a good physics reason for the origin of this divergence or why it must be regularized in this fashion, other than that it works.  It would be very interesting to further explore this in the future.

We proceed as if the kinetic terms were diagonal and see what awaits us. The $\zeta_3(0)$ computations are uncomplicated: \FloatBarrier
{\renewcommand{\arraystretch}{1.5}
\begin{table}[h] \centering \begin{tabular}{rclrcl}
$\zeta_{3,1,0}(0)$&$=$ &$-\frac{1}{180}$ & $\zeta_{3,-1,0}(0)$&$=$&$\frac{269}{180}$ \nonumber \\
\rowcolor{Gray} $\zeta_{3,0,1}(0)$&$=$&$-\frac{41}{60}$ & $\zeta_{3,1,2}(0)$&$=$&$-\frac{31}{36}$ \nonumber \\
$\zeta_{3,\mathrm{spins}}(0)$ &$=$ & $\frac{1}{180}$ & $\zeta_{3,\mathrm{even~spins}}(0)$&$=$&$\frac{1}{180}$ \nonumber \\
\rowcolor{Gray} $\zeta_{3,\mathrm{PM~spins}}(0)$&$=$&$\frac{1}{20}$ & $\zeta_{3,\mathrm{even~PM~spins}}(0)$&$=$&$-\frac{19}{30}$\end{tabular}\end{table}
} \FloatBarrier

These add to ensure that $\zeta_3(0)=0$ for all four theories in question. 

Now turn to $\zeta_3^\prime(0)$. There are two obstructions; both $K_{3,\mathrm{new~scalar}}$ and $K_{3,\mathrm{new~tensor}}$ are divergent/ill-defined, arising from precisely the two particles we expected subtlety from. First, we state results for everything else, then turn our attention to the obstructions. \FloatBarrier
{\renewcommand{\arraystretch}{1.5}
\begin{table}[!htbp]
\centering
\begin{tabular}{rcl}
$J_{3,\mathrm{scalar}}$&$=$&$-\frac{\log (A)}{24}+\frac{7 \zeta '(4)}{32 \pi ^4}-\frac{7 \gamma }{2880}+\frac{53}{4320}+\frac{\log (4096)}{8640}-\frac{7 \log (\pi )}{2880} $\nonumber \\
\rowcolor{Gray} $K_{3,\mathrm{scalar}}$&$=$&$ -\frac{\log (A)}{8}+\frac{5}{8} \zeta '(-3)-\frac{\zeta (3)}{8 \pi ^2}+\frac{11}{1152}-\frac{11 \log (2)}{2880} $\nonumber \\
$J_{3,\mathrm{new~scalar}}$&$=$&$  -\frac{\log (A)}{24}+\frac{7 \zeta '(4)}{32 \pi ^4}-\frac{7 \gamma }{2880}+\frac{7253}{4320}+\frac{\log (4096)}{8640}-\frac{7 \log (\pi )}{2880} $\nonumber \\
\rowcolor{Gray} $J_{3,\mathrm{new~vector}}$&$=$&$ -\frac{9 \log (A)}{8}+\frac{21 \zeta '(4)}{32 \pi ^4}-\frac{7 \gamma }{960}+\frac{1133}{1440}+\frac{7 \log (2)}{80}-\frac{7 \log (\pi )}{960} $\nonumber \\
$K_{3,\mathrm{new~vector}}$&$=$&$ -\frac{27 \log (A)}{8}+\frac{15}{8} \zeta '(-3)-\frac{9 \zeta (3)}{8 \pi ^2}+\frac{57}{128}+i \pi -\frac{91 \log (2)}{960}$ \nonumber \\
\rowcolor{Gray} $J_{3,\mathrm{new~tensor}}$&$=$&$ -\frac{125 \log (A)}{24}+\frac{35 \zeta '(4)}{32 \pi ^4}-\frac{7 \gamma }{576}+\frac{413}{864}+\frac{61 \log (2)}{144}-\frac{7 \log (\pi )}{576} $\end{tabular}\end{table}} 
\FloatBarrier
{ Note the imaginary part $i\pi$ in $K_{3,\mathrm{new~vector}}$.} Now the spin sums:
{\renewcommand{\arraystretch}{1.5}
\begin{table}[!htbp]
\centering
\begin{tabular}{rcl}
$J_{3,\mathrm{spins}}$&$=$&$\frac{\log (A)}{24}-\frac{7 \zeta '(4)}{32 \pi ^4}-\frac{53}{4320}+\frac{7 \gamma }{2880}-\frac{\log (2)}{720}+\frac{7 \log (\pi )}{2880} $\nonumber \\
\rowcolor{Gray} $K_{3,\mathrm{spins}}$&$=$&$ \frac{\log (A)}{8}-\frac{5}{8} \zeta '(-3)+\frac{\zeta (3)}{8 \pi ^2}-\frac{11}{1152}+\frac{11 \log (2)}{2880} $\nonumber \\
$J_{3,\mathrm{even~spins}}$&$=$&$\frac{\log (A)}{24}-\frac{7 \zeta '(4)}{32 \pi ^4}-\frac{53}{4320}+\frac{7 \gamma }{2880}-\frac{\log (2)}{720}+\frac{7 \log (\pi )}{2880} $\nonumber \\
\rowcolor{Gray} $K_{3,\mathrm{even~spins}}$&$=$&$ \frac{\log (A)}{8}-\frac{5}{8} \zeta '(-3)+\frac{\zeta (3)}{2 \pi ^2}-\frac{11}{1152}-\frac{709 \log (2)}{2880} $\nonumber \\
$J_{3,\mathrm{PM~spins}}$&$=$&$ \frac{51 \log (A)}{8}-\frac{63 \zeta '(4)}{32 \pi ^4}-\frac{471}{160}+\frac{7 \gamma }{320}-\frac{1}{80} 41 \log (2)+\frac{7 \log (\pi )}{320}$ \nonumber \\
\rowcolor{Gray} $K_{3,\mathrm{PM~spins}}$&$=$&$ -\frac{39 \log (A)}{8}-\frac{45}{8} \zeta '(-3)+\frac{19 \zeta (3)}{8 \pi ^2}+\frac{1823}{384}+\frac{171 \log (2)}{320}$\nonumber \\
$J_{3,\mathrm{even~PM~spins}}$&$=$&$ \frac{21 \log (A)}{4}-\frac{21 \zeta '(4)}{16 \pi ^4}-\frac{1553}{720}+\frac{7 \gamma }{480}-\frac{1}{40} 17 \log (2)+\frac{7 \log (\pi )}{480} $\nonumber \\
\rowcolor{Gray} $K_{3,\mathrm{even~PM~spins}}$&$=$&$ -\frac{33 \log (A)}{4}-\frac{15}{4} \zeta '(-3)+\frac{13 \zeta (3)}{8 \pi ^2}+\frac{997}{192}-\frac{1}{480} 869 \log (2)$\end{tabular}\end{table}} 
\FloatBarrier
We now turn to $K_{3,\mathrm{new~scalar}}$. This $K$ function involves the following integral:
\begin{equation}K_{3,\mathrm{new~scalar}}^{\mathrm{div}}=\frac{1}{12}\int_0^{-\frac{5}{2}}dx~x(1-4x^2)\psi\left(\frac{1}{2}+x\right)\, .\end{equation}
The integral does not converge as is, so we shift the upper region of integration to $-\frac{5}{2}+\epsilon$. Then, we may perform the integral, and expand the resulting answer in powers of $\epsilon$.  The terms which survive as $\epsilon\rightarrow 0$ are
\begin{equation}K_{3,\mathrm{new~scalar}}^{\epsilon}=-\frac{49 \log (A)}{8}-5 \ln \epsilon+\frac{5}{8} \zeta '(-3)-\frac{5 \zeta (3)}{8 \pi ^2}-\frac{6037}{1152}-i \pi -\frac{7211 \log (2)}{2880}-\frac{5 \log (\pi )}{2}\, .\end{equation}
We see that this diverges logarithmically as $\epsilon\rightarrow 0 $. The same is true of the spin two; its $K$ is associated with the integral
\begin{equation}K_{3,\mathrm{new~tensor}}^{\mathrm{div}}=\frac{125}{12}\int_0^{-\frac{1}{2}} dx~x\left(1-\frac{4}{25}x^2\right) \psi\left(\frac{1}{2}+x\right)\, .\end{equation}
Again, we may deform the limit of the integral to $-\frac{1}{2}+\epsilon$ (with the same $\epsilon$) and expand, keeping terms which survive as $\epsilon\rightarrow 0$:
\begin{equation}K_{3,\mathrm{new~tensor}}^{\epsilon}=\frac{115 \log (A)}{8}+5 \ln \epsilon+\frac{25}{8} \zeta '(-3)-\frac{5 \zeta (3)}{8 \pi ^2}+\frac{55}{1152}+\frac{1189 \log (2)}{576}+\frac{5 \log (\pi )}{2}\end{equation}
We see that upon adding these two together, the divergences in $\epsilon$ cancel and we may take a smooth $\epsilon \rightarrow 0$ limit, obtaining a finite result,
\begin{equation}
K_{3,\mathrm{new~scalar~and~tensor}}=\frac{33 \log (A)}{4}+\frac{15}{4} \zeta '(-3)-\frac{5 \zeta (3)}{4 \pi ^2}-\frac{997}{192}-i \pi -\frac{211 \log (2)}{480}\,.\end{equation}

Note that if, instead, we had regularized the scalar by $\epsilon$ and the tensor by $2\epsilon$, then the divergences would still have cancelled, but the answer would have differed by $5\ln 2$, which would not give the expected result, as we will show below. Again, we do not yet have a good motivation for using the same $\epsilon$ for both, other than that it gives the expected answers.

Now, we're ready to put the pieces together. First the nonminimal Vasiliev theory:
\begin{align}-\frac{1}{2}\zeta^{\mathrm{nonmin}\prime}_{hs,3}(0) &= -\frac{1}{2}\Big(J_{3,\mathrm{scalar}}+K_{3,\mathrm{scalar}}+J_{3,\mathrm{spins}}+K_{3,\mathrm{spins}}\Big) \nonumber \\
&=0\, .\end{align}
Then the minimal Vasiliev theory:
\begin{align}-\frac{1}{2}\zeta^{\mathrm{min}\prime}_{hs,3}(0) &= -\frac{1}{2}\Big(J_{3,\mathrm{scalar}}+K_{3,\mathrm{scalar}}+J_{3,\mathrm{even~spins}}+K_{3,\mathrm{even~spins}}\Big) \nonumber \\
&=\frac{\log (4)}{16}-\frac{3 \zeta (3)}{16 \pi ^2}\, .\end{align}
This is the free energy of a real $\square$ scalar on $S^3$. Now, the nonminimal PM theory:
\begin{align}-\frac{1}{2}\zeta^{\mathrm{nonmin}\prime}_{hs_2,3}(0) &= -\frac{1}{2}\Big(J_{3,\mathrm{scalar}}+K_{3,\mathrm{scalar}}+J_{3,\mathrm{new~scalar}}+J_{3,\mathrm{new~vector}}+K_{3,\mathrm{new~vector}}+J_{3,\mathrm{new~tensor}} \nonumber \\
&\qquad \qquad +K_{3,\mathrm{new~scalar~and~tensor}}+J_{3,\mathrm{spins}}+K_{3,\mathrm{spins}}+J_{3,\mathrm{PM~spins}}+K_{3,\mathrm{PM~spins}}\Big) \nonumber \\
&=0\, ,\end{align}
and finally, the minimal PM theory:
\begin{align}-\frac{1}{2}\zeta^{\mathrm{min}\prime}_{hs_2,3}(0) &= -\frac{1}{2}\Big(J_{3,\mathrm{scalar}}+K_{3,\mathrm{scalar}}+J_{3,\mathrm{new~scalar}}+J_{3,\mathrm{new~tensor}}+K_{3,\mathrm{new~scalar~and~tensor}} \nonumber \\
&\qquad \qquad +J_{3,\mathrm{even~spins}}+K_{3,\mathrm{even~spins}}+J_{3,\mathrm{even~PM~spins}}+K_{3,\mathrm{even~PM~spins}}\Big) \nonumber \\
&=-\frac{3 \zeta (3)}{8 \pi ^2}+\frac{i \pi }{2}+\frac{\log (1024)}{8} \, \\
&\approx 0.820761\, +1.5708 i\, .
\end{align}
As we demonstrated in \cite{Brust:2016gjy}, this is the free energy of a real $\square^2$ scalar on $S^3$, albeit in a different and simpler form than we presented there.

It would be interesting to understand more deeply the $\epsilon$-regulation that we do to obtain a finite result, beginning from the mixed ${\rm AdS}_4$ scalar-tensor theory we describe in \cite{ustoappear}. As we mentioned above, we were motivated by cancellation of the divergence, rather than any deep physics reason for why we should regulate in precisely this fashion.  It would be much more appealing if we had a natural  motivation for regularizing in the manner that we did. Such divergences have appeared before \cite{Giombi:2013yva}, although there, they signalled the presence of $\ln N$ corrections to $G_N^{-1}$. Nevertheless, perhaps a connection could be made to the induced gauge symmetries of their work.

It would also be interesting to obtain a deeper understanding of the imaginary piece that appears in the free energy, both from an AdS perspective as well as a CFT one. In the CFT, we can see that in three dimensions, the spectrum of $D$ (and therefore the AdS spectrum of $H$) is unbounded from below, due to the presence of the operator $j_0^{(0)}$ with $\Delta < 0$. This is unlike the other dimensions we study in this paper, and may be related to the presence of a complex free energy.


\section{Conclusions}
\label{sec:conclusions}

We have computed the one-loop partition functions in a generalization of Vasiliev's theory which includes a tower of partially massless modes.  By AdS/CFT, this theory is dual to a $U(N)$ or $O(N)$ free scalar CFT with a $\square^2$ kinetic term, and the bulk partition function we compute should match the sphere partition function of the $\square^2$ CFT.

We computed the one-loop partition functions by evaluating zeta functions for each particle within the theory, then summing up over each tower of spins.  The sum over spins requires additional regularization which must be compatible with the symmetries of the theory.

We computed the one-loop effective action for both the ``minimal'' version of the theory containing only even spins, and for the ``non-minimal'' version of the theory containing all spins. We did this in odd AdS dimensions $D=7$ through $19$, for which the log-divergent part of the effective action is dual to the $a$-type conformal anomaly of the dual boundary theory, and also for even-dimensional AdS spaces for $D=4$ through $8$, for which the finite part of the one-loop effective action is dual to the free energy on a sphere of the dual boundary theory. There were subtleties in the case $D=4$, but not $D=7$, the cases where module mixing occurs in the dual field theory. In $D=4$, there were divergences associated with $\zeta_3^\prime(0)$ for the new scalar and new tensor, which we were able to regulate by jointly shifting their masses. After regulating, the answer became finite and we could take the regulator to 0. However, we were forced to regulate in a particular way, using the same regulator for both particles, in order to obtain the expected results. We want to attempt to understand the motivation for this regularization in the future.

We found that in even $D$, in order to ensure that the finite part of the effective action is unambiguous (i.e. $\zeta_d(0)=0$), we needed to regulate the sum over partially massless spins by inserting $\left(s+\frac{d-5}{2}\right)^{-\alpha}$ before summing, sum, and then take the $\alpha \rightarrow 0 $ limit, just as in \cite{Gunaydin:2016amv}. 

Our results are that in the nonminimal theory, the one loop contribution vanishes and so there are no quantum corrections to the Newton's constant, and in the minimal theory the inverse Newton's constant gets a one-loop correction of exactly the same magnitude as in the original Vasiliev theory,
\be G_N^{-1}\propto \begin{cases} N \, ,& {\rm nonminimal/U(N)\ PM\ theory,}\\ N-1 \, ,& {\rm minimal/O(N)\ PM\ theory.} \end{cases}\ee
These results provide evidence that the theory is UV complete and that this computation is one-loop exact.

In the future, it would be interesting to understand better the nature of the divergences in the ${\rm AdS}_4$ case, and see if these sorts of subtlety occur in other theories besides the PM theories. It would also be interesting to attempt to explore the one-loop effective actions of PM theories dual to more general $\square^k$ theories. Also, we could explore the de Sitter analogue of this computation, or the adjoint scalar variant. Perhaps the fermionic $\slashed{\partial}^k$ theories' PM duals, or the supersymmetric extension of this theory, could shine some light on the puzzle related to the one-loop effective action in the type B Vasiliev theory, explored recently in \cite{Pang:2016ofv} and references therein. It would be also very interesting to attempt to explore the one-loop matching in non-integer $d$, computing directly $\tilde{F}$ at one-loop in the bulk and in the CFT. Finally, it would be interesting to explore the connection to the character-based approach taken in \cite{Bae:2016rgm}.

\acknowledgments{We thank Simone Giombi, Euihun Joung, Igor Klebanov, Ben Safdi and Evgeny Skvortsov for helpful discussions.
Research at Perimeter Institute is supported by the Government of Canada through Industry Canada and by the Province of Ontario through the Ministry of Economic Development and Innovation.
 
 \bibliographystyle{utphys}
\addcontentsline{toc}{section}{References}
\bibliography{adszeta}

\end{document}